\documentclass[authoryear,preprint,12pt]{elsarticle}
\journal{NeuroImage}

\usepackage[margin=2cm]{geometry}
\usepackage[T1]{fontenc}
\usepackage{float}
\usepackage[utf8]{inputenc}

\usepackage{charter}

\usepackage{color}
\usepackage[monochrome]{xcolor}

\usepackage{amssymb,amsmath,bm}
\DeclareMathOperator*{\argmax}{arg\,max}
\usepackage{lineno} 

\begin{document}
\begin{frontmatter}

\title{\textbf{A tutorial on generalized eigendecomposition for {\color{red}denoising, contrast enhancement, and dimension reduction} in multichannel electrophysiology}}

\author[inst1]{Michael X Cohen}

\affiliation[inst1]{organization={Donders Centre for Medical Neuroscience, Radboud University Medical Center, the Netherlands}}

\begin{abstract}
The goal of this paper is to present a theoretical and practical introduction to generalized eigendecomposition (GED), which is a robust and flexible framework used for dimension reduction and source separation in multichannel signal processing. In cognitive electrophysiology, GED is used to create spatial filters that maximize a researcher-specified contrast. {\color{red}For example, one may wish to exploit an assumption that different sources have different frequency content, or that sources vary in magnitude across experimental conditions.} GED is fast and easy to compute, performs well in simulated and real data, and is easily adaptable to a variety of specific research goals. This paper introduces GED in a way that ties together myriad individual publications and applications of GED in electrophysiology, and provides sample MATLAB and Python code that can be tested and adapted. Practical considerations and issues that often arise in applications are discussed. 
\vspace{3em}
\end{abstract}


\end{frontmatter}


\section{Background and motivation} \label{sec:sample1}

\subsection{What are "sources" and (how) can they be separated?}
The brain is an unfathomably complex and dynamic system, with countless neural operations that are characterized by both segregation and integration. Understanding the brain therefore requires a balance between isolating neurocognitive elements while maintaining sufficient ecological validity to generalize from experimental control to real-world behavior. One of the primary difficulties in neuroscience research is that myriad neural and cognitive operations unfold simultaneously, and can overlap in time, frequency, and space. In other words, separating sources of cognitive and neural processes is one of the major challenges in neuroscience research.

The term "source" can have several interpretations: It can refer to a single physical location in the brain, a distributed set of locations, a neural ensemble, a single neuron, a synapse, a cognitive operation, a computational algorithm that a neural ensemble produces, a neurochemical modulation, etc. Therefore, the mechanism of source separation depends on the goals of the research. For example, at a behavioral level, careful experiment design can be used to separate attentional from sensory sources of reaction time variability. When studying neural oscillations, we can consider the Fourier transform to be a method of spectral source separation under the assumption that sources are mixed in the time domain but have non-overlapping frequency characteristics. FMRI is highly suitable for spatial source separation due to the small voxel sizes and relative locality of the BOLD response.

The type of source separation under consideration in this paper is a descriptive-statistical separation, where sources are isolated based on spatiotemporal patterns in a channel covariance matrix (a covariance matrix contains the pairwise linear interactions across channels). These sources can stem from a spatially restricted spatial location (e.g., modeled by a single dipole) or they can stem from an anatomically distributed but synchronous network. As explained later, GED has no anatomical constraints, and the sources are defined according to descriptive-statistical criteria, in particular, the information contained in covariance matrices.

Neural data are noisy and variable, and there are more sources in the brain than measurement points. This means that applying a source separation method does not guarantee that a single source has been isolated. (The same uncertainty holds for other methods as well; for example, an fMRI study may indicate a localized source while many other sources were involved but below the statistical threshold.) In this sense, source separation methods should be seen as \textit{attempts} to isolate sources, and their accuracy depends on a variety of experimental and data-analytic factors. {\color{red}A more conservative interpretation of GED is that it acts as a contrast-enhancing filter by isolating relevant from irrelevant patterns in the data while simultaneously allowing for a reduced-dimensional signal to analyze.}

\subsection{Why multivariate analyses?}
It is not contentious to write that electrophysiologists are uninterested in \textit{electrodes}; instead, electrophysiologists are interested in using the numerical values that the electrodes provide to understand how the brain works. It is equally uncontentious to state that there is no one-to-one mapping of electrode to computational source. The neural computations underlying cognition are implemented by complex interactions across neural circuits comprising various types of cells, modulations by neurochemicals, etc.; these circuit interactions produce electromagnetic fields that can be quantified using LFP, EEG, or MEG (local field potential, electroencephalography, magnetoencephalogray), but these electrical fields propagate to multiple electrodes simultaneously. Furthermore, each electrode simultaneously measures the electrical fields from multiple distinct neural circuits, plus artifacts from muscles and heart beats, and from noise.

\begin{figure}[H] 
    \centering
    \includegraphics[width=.8\textwidth]{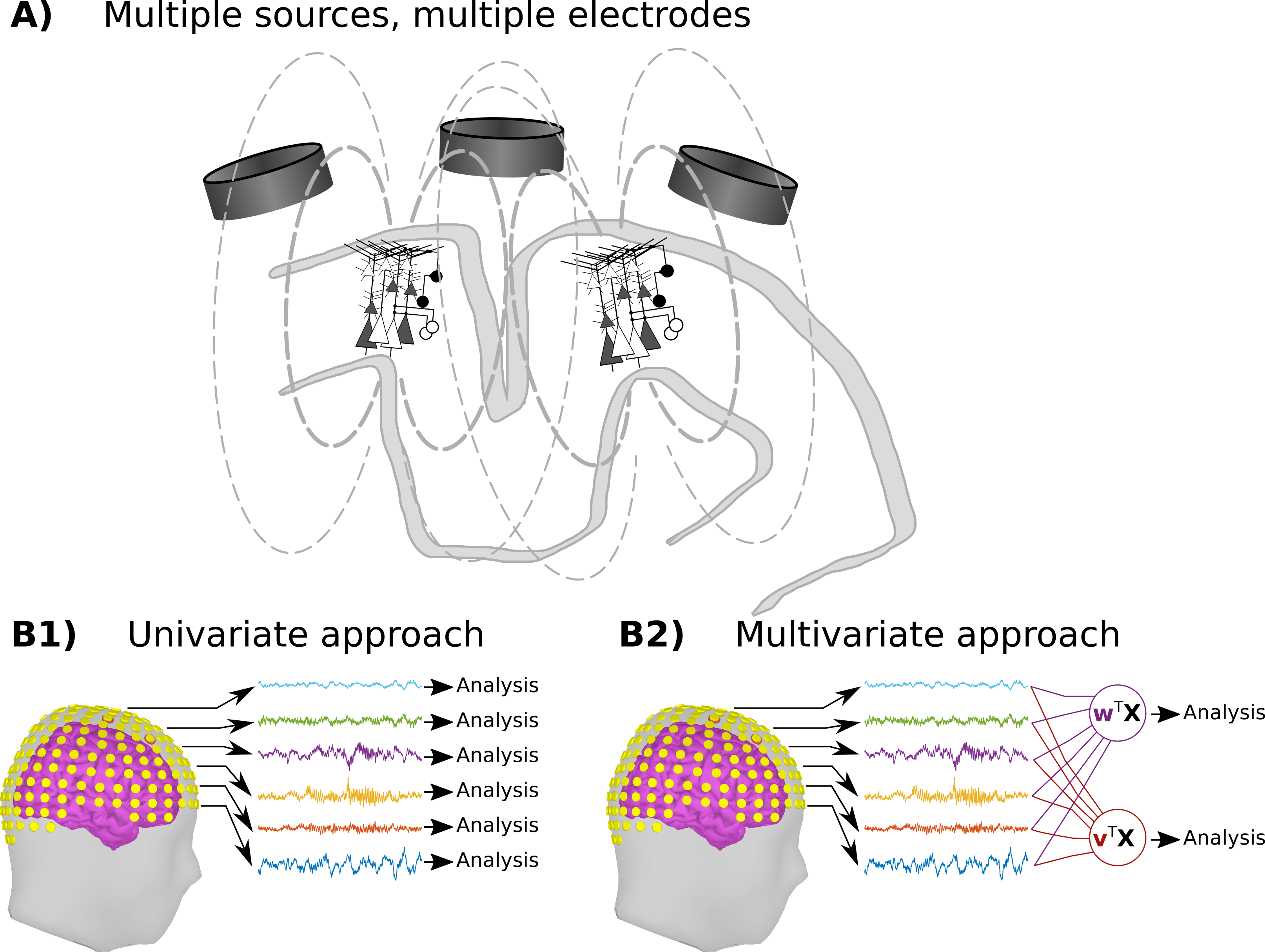}
    \caption{A) An anatomical interpretation of the source-separation problem. Electrophysiology involves measuring data from electrodes (black disks), which are manifest (observable) variables. But electrophysiologists are interested in understanding neural circuits (latent variables), which are complex combinations of neurons and neurochemicals, and which generate electrical fields that propagate to the electrodes. The challenge is to leverage the spatial correlations across electrodes to separate activity from different neural sources. The illustration here makes it seem like the problem is one of anatomical assignment, but because neural computations are spatially distributed, a single localized dipole is not physiologically plausible for all high-level sensory/cognitive/motor phenomena. Therefore, the goal of source separation can be treated as one of statistical assignment. B) The distinction between univariate (B1) and multivariate (B2) analyses lies in the conceptual and statistical use of the electrodes: The univariate approach treats each electrode as an independent measurement; the multivariate approach considers that the signals of interest are embedded in patterns that span many electrodes, and thus isolating those signals requires appropriately designed spatial filters (vectors $\mathbf{w}$ and $\mathbf{v}$ are spatial filters and $\mathbf{X}$ is the channels-by-time data matrix.} \label{fig:fig1}
\end{figure}

Thus, the manifest variables (a.k.a. observable data) that we measure (voltage fluctuations in electrodes) are indirect and comprise mixtures of the latent constructs we seek to understand. The mixing of sources at the electrodes motivates \textit{multivariate} analyses that identify patterns distributed across electrodes. This can be contrasted with \textit{univariate} analyses that consider each individual electrode to be a separate statistical unit (Figure \ref{fig:fig1}).

And yet, mass-univariate analyses have remained the most common analysis method for decades. The mass-univariate approach is nearly universally used because it is simple and has been successful throughout the history of neuroscience, and because electrode technology and computational power limited the types of analyses that were feasible in the past. However, these limitations no longer exist, and thus there is little justification for maintaining the position that analyses that dominated the literature decades ago are \textit{prima facie} optimal today.

Thus, the argument here is not that mass-univariate analyses are wrong or misleading; rather, the argument is that progress in neuroscience will be accelerated by shifting to conceptualizing and analyzing data with the goal of isolating and extracting information that is distributed across a set of electrodes. This argument has been made several times before \citep{Makeig2002Jan,Uhlhaas2009Jul,Hebb1949,Buzsaki2010Nov}, although the increase in simultaneously recorded measurements, computational power, and analysis possibilities makes multivariate methods more attractive, feasible, and insightful now than in previous decades.

(Although the focus of this paper is on electrophysiology, the previous argument applies to any set of brain or behavioral manifest variables, including neurons, fMRI voxels, two-photon imaging pixels, reaction time, questionnaire items, etc.)

To be sure, there are myriad modern data analysis methods that leverage increased understanding of neurophysiology and computational power. Here I will focus on one family of spatial multivariate analyses, which are used as dimension-reducing, contrast-enhancing spatial filters. A spatial filter is a set of weights, such that the weighted combination of the manifest variables optimizes some criteria. That weighted combination results in a reduced number of virtual data channels (often called "components") compared to the original dataset, and that are designed such that each component isolates a pattern in the data that may be mixed or have low signal-to-noise quality in individual electrodes.

The purpose of this tutorial paper is to introduce a family of methods based on generalized eigendecomposition (GED). The GED seed can sprout many seemingly different multivariate applications, which makes it a powerful method to adapt to specific hypotheses, research goals, and dataset types. There are several other excellent general introductions to GED in the neuroscience literature \citep{Parra2005Nov,de2014joint,haufe2014interpretation}; the focus of the present paper is on practical issues that researchers might face when implementing GED.

\subsection{A multitude of multivariate analyses}
There are many multivariate analysis methods that have been introduced and validated on electrophysiology data. Below is a brief description of several commonly used methods. I focus on their limitations only to facilitate a contrast with the advantages of GED, without implying that these methods are flawed or should be avoided.

Principal components analysis (PCA) is a popular dimension-reduction method that finds a set of channel weights, such that the weighted combination of channels maximizes variance while keeping all components mutually orthogonal. {\color{red}PCA is an excellent tool for data compression, but has three limitations with regards to contrast enhancement, denoising, and source separation}: It is descriptive as opposed to inferential; the PC vectors are constrained to be orthogonal in the channel space (which means that correlated sources in channel space remain correlated in PCA space); and maximizing variance does not necessarily maximize relevance \citep{de2014joint}.

Independent components analysis (ICA) is nearly ubiquitously used in M/EEG research to attenuate artifacts such as eye blinks and muscle activity, and is also used for data analysis \citep{Debener}. ICA is a blind separation method that relies on the assumption that sources are {\color{red}statistically independent} and non-Gaussian distributed. It is often used as a descriptive measure, but cross-validation methods exist for evaluating statistical significance of individual components \citep{hyvarinen2011testing}. In simulated EEG data, ICA has low accuracy at recovering ground-truth sources \citep{cohen2017comparison,zuure2021narrowband}.

Decoding, a.k.a. multivariate pattern analysis, involves using machine-learning methods to classify experiment conditions (e.g., a particular motor response or visual stimulus) based on weighted combinations of brain signals \citep{Cichy2017Sep,King2014Apr}. Some linear classifiers (e.g., Fisher linear discriminant analysis) are built on GED, however, decoding methods generally threshold and binarize the data, thus discarding a considerable amount of rich and meaningful spectral and temporal variability.

Deep learning is a framework for mapping inputs to outputs, via myriad simple computational units, each of which implements a weighted sum of its inputs plus a nonlinearity \citep{Schirrmeister2017Nov}. Deep learning has been transformative in many computational fields including computer vision and language translation. Outside of the visual system, deep learning applications in neuroscience have not (yet) made a major impact. Part of the difficulty of deep learning in neuroscience is that its representations are complex, nonlinear, and difficult to interpret. In other words, deep learning has major applications in society and engineering, but is (so far) of limited value for providing mechanistic insights (the same can be said of decoders more generally) \citep{Ritchie2019Jun}. On the other hand, deep learning and linear decompositions could be used synergistically, for example by using linear decompositions to reduce dimensionality and enhance the signal-to-noise characteristics of data that are categorized by a deep learning network \citep{Lawhern2018Oct}.

There are several other multivariate components analyses that are less commonly used in neuroscience, such as factor analysis, Tucker decomposition, and nonnegative matrix factorization. An exhaustive list and comparison of multivariate components analyses is beyond the scope of this paper.

\subsection{Motivation for and advantages of GED}
{\color{red}GED as a tool for denoising, dimension reduction, and source separation} of multichannel data has several advantages. First, it is based on specifying hypotheses. A cornerstone of experimental science is generating and testing null and alternative hypotheses. This manifests in statistical comparisons as a difference or a ratio between the means of two sample distributions. The spatial filters created by GED are designed to maximize a contrast between two features of the data --- a feature to enhance, and a feature that acts as reference. Examples of feature-pairs include: an experiment condition and a control condition; a prestimulus period and a poststimulus period; the trial-average and the single-trial data; narrowband filtered and unfiltered data. When the two covariance matrices computed from those data features are equivalent, the GED returns a contrast ratio of 1, which is the expected null-hypothesis value. In this sense, GED is a supervised method, which can be contrasted with unsupervised methods like PCA or ICA.

Second, because of the inherent comparison of two covariance matrices, GED allows for inferential statistics to determine whether a component is significant. Methods for statistical inference are described in a later section.

Third, GED has only a few key researcher-guided analysis choices, which makes it easy to learn, apply, and adapt to new situations.

Fourth, and related to the previous point, there are no spatial or anatomical constraints. The order and relative locations of the physical data channels (electrodes, sensors, pixels, or voxels) does not affect the analysis. This means that the spatial maps can be physiologically interpreted without concern for biases imposed by an \textit{a priori} anatomical or physical model.

Fifth, GED allows for individual differences in topographies. For example, alpha-band activity might be maximal at electrodes Pz, POz, Oz, PO7, etc., in different individuals, leading to possible difficulties and subjectivity with electrode selection. A GED that maximizes alpha-band activity allows for idiosyncratic functional-anatomical distributions for different individuals, while ensuring that the components across all individuals satisfy the same statistical criteria.

Sixth, GED is deterministic and non-iterative. This means that repeated decompositions of the same data give the same solution (this can be contrasted, for example, with ICA algorithms that are initialized with random weights). And it means that GED is fast. Indeed, the GED typically takes a few milliseconds to compute; most of the total analysis time comes from data preparation such as temporal filtering.

Finally, GED has a long history of applications in statistics, machine learning, engineering, and signal processing \citep{Parra2005Nov}. Although not always called "GED," generalized eigendecomposition provides the mathematical underpinning of many analysis methods, including linear discriminant analysis, common-spatial pattern (used in brain-computer interface algorithms), blind-source separation \citep{tome2006generalized,blankertz2007optimizing}, and other methods discussed later. Although there are "tricks" for optimizing GED for specific applications, the general approach is well established in multiple areas of science \citep{parra2003blind,Sarela2005Mar}.

\subsection{GED in the wild}
GED is widely used in neuroscience, although the terminology differs. In brain-computer interfaces, GED is called common spatial pattern analysis \citep{blankertz2007optimizing} and is used to design spatial filters that facilitate neural control over a computer program \citep{Rivet2011Aug,Ai2018Jan}. \cite{nikulin2011novel} used GED to design a narrowband spatial filter, which they termed spatio-spectral decomposition. This method was extended to use broadband energy compared to narrowband energy over a successive range of center frequencies, and was termed spectral scanning \citep{de2015scanning} or, more generally, joint decorrelation \citep{de2014joint}. \cite{dahne2014spoc} used GED to design a spatial filter that maximizes a correlation between EEG and a behavioral measure like reaction time. \cite{cohen2017multivariate} adapted GED to identify multivariate cross-frequency coupling. Several groups have used GED to optimize ERPs, which is particularly useful for single-trial analyses \citep{rivet2013optimal,Tanaka2019Aug,Das2020Jan}. GED has been used to obtain components that have maximal signal-to-noise characteristics in steady-state evoked potential studies \citep{Dmochowski2015Apr,Cohen2017RESS}. GED may also provide an efficient method for spike-sorting in invasive multichannel recordings \citep{Wouters2018Oct}, as well as online epileptic spike waveform detection \citep{Dan2020Nov}.

There are many other applications of GED in electrophysiology research; the goal here is not to cite all of them, but instead to highlight that GED is a core part of the corpus of multivariate neuroscience analyses, even if its inclusion is not apparent from the titles of research papers.

\subsection{Brief overview of GED}
\begin{figure}[H]
    \centering
    \includegraphics[width=.8\textwidth]{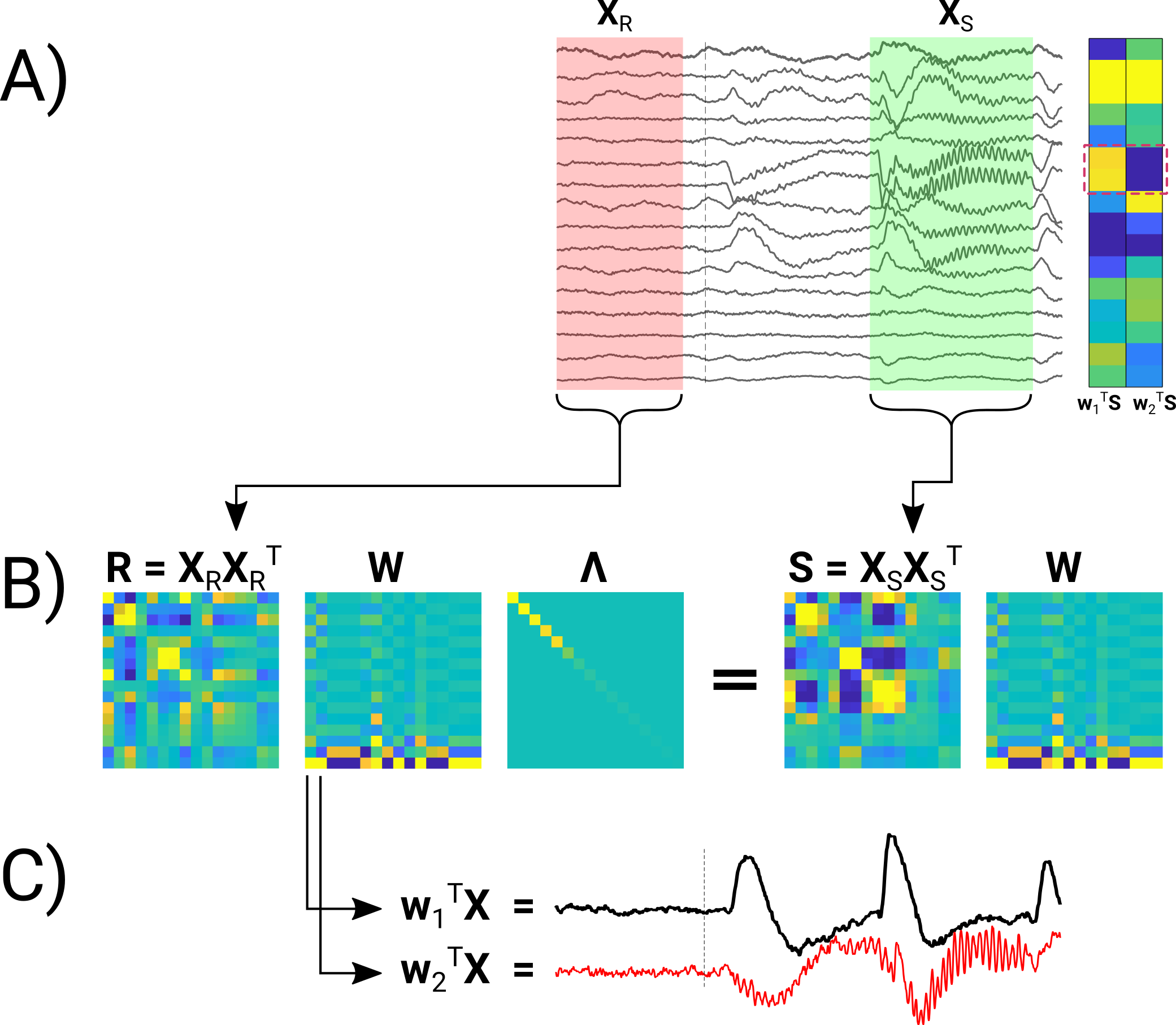}
    \caption{Graphical overview of GED. This example shows laminar recordings in mouse V1 after a visual stimulus onset. A) Two time windows are selected for the "reference" ($\mathbf{X_R}$) and "signal" ($\mathbf{X_S}$) features of the data. B) A generalized eigenvalue decomposition is performed on the two covariance matrices created from the two time windows. The resulting eigenvectors (matrix $\mathbf{W}$) are the spatial filters, and their corresponding eigenvalues (diagonal elements of matrix $\mathbf{\Lambda}$) encode the ratio of matrix $\mathbf{S}$ to $\mathbf{R}$ along each direction $\mathbf{w_i}$. C) Each eigenvector times the data produces a component, with an accompanying time series and spatial map (spatial maps are visualized to the right of the time series in panel A). In this example, the first component captured a low-frequency response to the visual stimulus, while the second component isolated the gamma-band response. The red dashed box in the spatial maps indicates the approximate location of cortical layer 4.} \label{fig:overview}
\end{figure}

Before delving into the details, it will be useful to have the "bird's eye view" of GED (Figure \ref{fig:overview}). GED is a decomposition of two covariance matrices, here termed $\mathbf{S}$ and $\mathbf{R}$. These two covariance matrices come from different features of the data: an experiment condition and a control condition, narrowband filtered and unfiltered data, etc. The $\mathbf{S}$ matrix is the covariance of the "signal" data feature of interest, and the $\mathbf{R}$ matrix is the covariance of the "reference" data that provides a comparison.

The GED finds a weighting of the data channels that maximizes a signal-to-noise ratio (SNR) that can be thought of as $\mathbf{S}/\mathbf{R}$ (division is not defined for matrices, but this conceptualization can be helpful). Inter-channel covariance patterns that are common between $\mathbf{S}$ and $\mathbf{R}$ are ignored.

The channel weight vector associated with the largest eigenvalue is the spatial filter, and the weighted sum of all channel time series is the component time series that maximizes the researcher-defined criteria established by selecting data for the two covariance matrices. That component time series can be used in standard analyses such as ERP or time-frequency analysis, and its accompanying topography can be visualized for topographical or anatomical interpretation.

\begin{figure}[H]
    \centering
    \includegraphics[width=.8\textwidth]{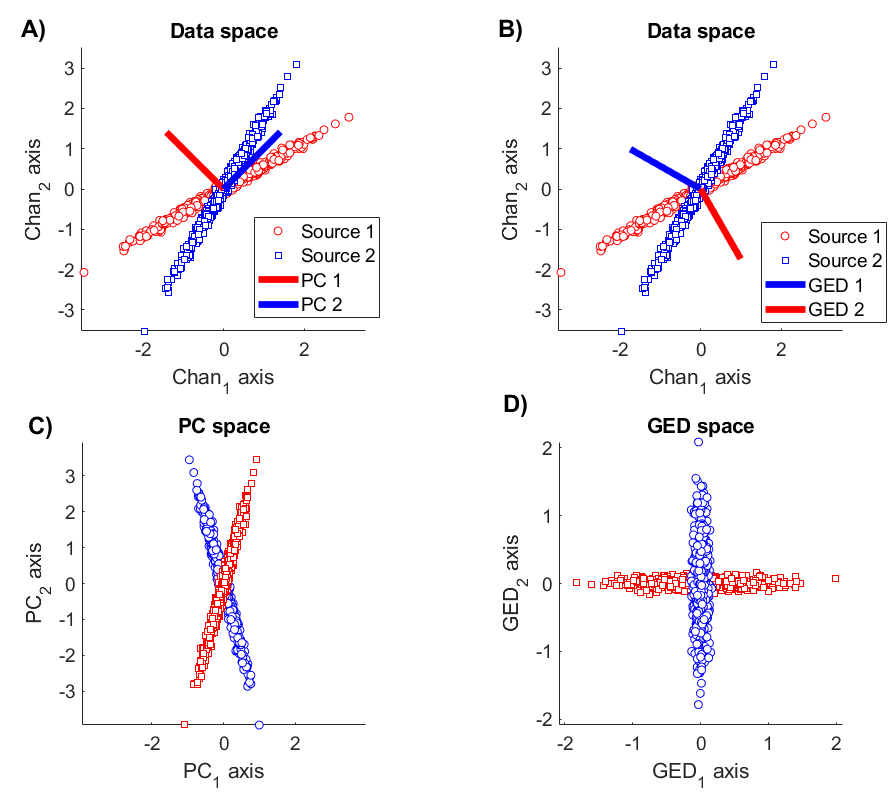}
    \caption{Simple example comparing PCA and GED. (A, B) The data were simulated as two "streams," indicated by the blue and red markers. Both plots show the same data. (C) A PCA on these data removes the correlation between variables, but does not separate the sources. The PC basis vectors, which form the axes PC$_1$ and PC$_2$, are shown in the channel space in panel A. (D) A GED on the channel covariance matrices for the red vs. blue streams was successful at source separation. The GED eigenvectors, which form axes GED$_1$ and GED$_2$, are shown in the channel space in panel B. Notice that the sources are correlated in the data space and in the PC space, and are orthogonal in the GED space.} \label{fig:simpleExample}
\end{figure}

Figure \ref{fig:simpleExample} shows an example comparison of PCA and GED in a simulated 2D dataset. This illustration highlights several key features of GED, including the ability to separate sources that are correlated in the original channel space.

\section{Mathematical and statistical aspects}

\subsection{The math of GED}
The goal of GED is to find a set of channel weights that multiply the channel time series, such that the weighted sum of channels maximizes a contrast between two data features. For example, consider that the to-be-maximized feature is a post-stimulus time window, and that the pre-stimulus time window is the reference period. If the post-stimulus data are contained in a channels-by-time matrix $\mathbf{X}_S$, the pre-stimulus data are contained in a channels-by-time matrix $\mathbf{X}_R$, and the set of weights are contained in column vector $\mathbf{w}$, then the goal of GED can be expressed as:
\begin{equation} \label{eq:mainPowRatio}
    \lambda = \frac{\|\mathbf{w}^\text{T}\mathbf{X}_S\|^2}{\|\mathbf{w}^\text{T}\mathbf{X}_R\|^2}
\end{equation}
$\|\cdot\|^2$ indicates the squared magnitude of the vector (the sum of its squared elements). Thus, $\lambda$ is the ratio of the magnitude of the "signal" data filtered through $\mathbf{w}$, to the magnitude of the "reference" data filtered through the same $\mathbf{w}$. The squared magnitude of a vector can be expressed as the dot product of the row vector with itself. Thus:
\begin{align}
    \lambda &= \frac{(\mathbf{w}^\text{T}\mathbf{X}_S)(\mathbf{w}^\text{T}\mathbf{X}_S)^\text{T}}{(\mathbf{w}^\text{T}\mathbf{X}_R)(\mathbf{w}^\text{T}\mathbf{X}_R)^\text{T}}\\[1em]
     &= \frac{\mathbf{w}^\text{T}\mathbf{X}_S\mathbf{X}_S^\text{T}\mathbf{w}}{\mathbf{w}^\text{T}\mathbf{X}_R\mathbf{X}_R^\text{T}\mathbf{w}}\\[1em]
     &= \frac{\mathbf{w}^\text{T}(\mathbf{X}_S\mathbf{X}_S^\text{T})\mathbf{w}}{\mathbf{w}^\text{T}(\mathbf{X}_R\mathbf{X}_R^\text{T})\mathbf{w}}\label{eq:powerRatioParens}
\end{align}
The parentheses in equation \ref{eq:powerRatioParens} reveal that the solution to equation \ref{eq:mainPowRatio} can be developed through covariance matrices.

A covariance matrix is an $M\times M$ matrix in which the element in row $i$ and column $j$ contains the covariance between channels $M_i$ and $M_j$, defined as the sum of the element-wise multiplications of the mean-centered channel time series. Thus, the covariance matrix contains all linear pairwise interactions. A covariance is simply a non-normalized Pearson correlation coefficient, and thus a covariance matrix is a correlation matrix that retains the scale of the data (e.g., $\mu V$).

If the data are organized in a channels-by-time matrix $\mathbf{X}$ with $n$ time points, then the covariance matrix is given by
\begin{align} \label{eq:covariance}
    \mathbf{C} = \mathbf{XX}^\text{T}(n-1)^{-1}
\end{align}
In general, the size of the covariance matrix should be channels-by-channels, and so the multiplication would be expressed as $\mathbf{X}^\text{T}\mathbf{X}$ if the data were organized as time-by-channels. The division by $n-1$ is a normalization factor that prevents the covariance from increasing simply by increasing the number of observations (time points).

The data must be mean-centered before the multiplication. Mean-offsets in the data will cause the GED solutions to point in the direction of the offsets instead of the direction that maximizes the desired optimization criterion. Mean-centering means that the average value of each channel, over the time window used to compute the covariance matrix, is zero. 

Variance normalization is not necessary if all channels are in the same scale (e.g,. microvolts). Multimodal data might need within-modality normalization, which is discussed in section \ref{sec:multimodal}.

MATLAB code for computing the covariance matrix follows (variable \texttt{data} is channels-by-time).
\begin{verbatim}
data = data-mean(data,2); % mean-center
S = data*data' / (size(data,2)-1);
\end{verbatim}

As introduced in the previous section, GED requires two covariance matrices: the $\mathbf{S}$ matrix computed from features of the data to highlight, and the $\mathbf{R}$ matrix computed from features of the data to use as reference ($\mathbf{S}$ for \textit{signal} and $\mathbf{R}$ for \textit{reference}). The signal and reference data should be matched on as many features as possible, similar to how a control condition in an experiment should be matched to the experimental condition. In this sense, the GED can be thought of as maximizing a signal-to-\textit{reference} ratio instead of a signal-to-\textit{noise} ratio.

If $\mathbf{S}$ is computed from a time window during stimulus presentation and $\mathbf{R}$ is computed from a time window before stimulus presentation (the inter-trial interval), then the resulting GED will isolate the covariance patterns that maximally differentiate stimulus processing compared to the inter-trial-interval, while excluding any patterns in the data that are present before and during the stimulus period, such as ongoing spontaneous activity or slower neurocognitive processes that are unrelated to stimulus processing.

Once the two covariance matrices are formed, the goal is to find an $M$-element vector of weights (called vector $\mathbf{w}$; each element $w_i$ is the weight for the $i^{th}$ data channel), which acts as a \textit{spatial filter} that reduces the dimensionality of the data from $M$ channels to 1 component. The elements in $\mathbf{w}$ are constructed such that the linear weighted sum over all channels maximizes the ratio of "multivariate power ratio" in $\mathbf{S}$ to that in $\mathbf{R}$. Rewriting equation \ref{eq:powerRatioParens} using covariance matrices leads to the form $\mathbf{w}^\text{T}\mathbf{Sw}$, which is known as the quadratic form. The quadratic form is a single number that encodes the variance in matrix $\mathbf{S}$ along direction $\mathbf{w}$. Therefore, the goal of GED is to maximize the ratio of the quadratic forms of the two matrices, where that ratio is encoded in $\lambda$. 
\begin{equation} \label{eq:rayleigh}
    \lambda = \frac{\mathbf{w}^\text{T}\mathbf{Sw}}{\mathbf{w}^\text{T}\mathbf{Rw}}
\end{equation}
This expression is also known as the generalized Rayleigh quotient. Note that when the data covariance matrices are the same, $\lambda=1$, which can be considered the null-hypothesis value ($\text{H}_0: \mathbf{S}=\mathbf{R}$).

The goal of GED is to find the vector $\mathbf{w}$ that maximizes $\lambda$. This is the objective function.
\begin{equation}\label{eq:objectiveFun}
    \argmax_{\mathbf{w}} \frac{\mathbf{w}^\text{T}\mathbf{Sw}}{\mathbf{w}^\text{T}\mathbf{Rw}}
\end{equation}
Equation \ref{eq:objectiveFun} would produce only one spatial filter, however, it can be expanded to include additional vectors $\mathbf{w}_2$ up through $\mathbf{w}_M$, with each vector $\mathbf{w}_i$ subject to the constraint that it maximizes $\lambda_i$ while being uncorrelated with previous components (that is, the components are orthogonal in the GED space, as shown in Figure 3). {\color{red}It can be shown that} the expansion of one vector to a set of $M$ vectors means that equation \ref{eq:rayleigh} can be rewritten as follows.
\begin{equation}
    \mathbf{\Lambda} = (\mathbf{W}^\text{T}\mathbf{RW})^{-1}\mathbf{W}^\text{T}\mathbf{SW}
\end{equation}
where each column of $\mathbf{W}$ is a spatial filter, and each diagonal element of $\mathbf{\Lambda}$ is the multivariate ratio in the direction of the corresponding column in $\mathbf{W}$ \citep[additional details on the derivation of this solution can be found elsewhere, e.g. ][]{Ghojogh2019Mar,parra2003blind}. Some algebraic manipulations brings us to the matrix solution to the optimization problem:
\begin{align}
    \mathbf{\Lambda} &= (\mathbf{W}^\text{T}\mathbf{RW})^{-1}\mathbf{W}^\text{T}\mathbf{SW}\\
    \mathbf{\Lambda} &= \mathbf{W}^{-1}\mathbf{R}^{-1}\mathbf{W}^{-\text{T}}\mathbf{W}^\text{T}\mathbf{SW}\\
    \mathbf{\Lambda} &= \mathbf{W}^{-1}\mathbf{R}^{-1}\mathbf{SW}\\
    \mathbf{RW}\mathbf{\Lambda} &= \mathbf{SW} \label{eq:GEDmatrices}
\end{align}
Equation \ref{eq:GEDmatrices} is known as a generalized eigendecomposition on matrices $\mathbf{S}$ and $\mathbf{R}$. This means that the set of weights that maximizes the multivariate signal-to-noise ratio --- the spatial filter --- is an eigenvector, and the value of that ratio is the corresponding eigenvalue. It is also useful to see that equation \ref{eq:GEDmatrices} is conceptually equivalent to a "regular" eigendecomposition on a matrix product.
\begin{align}
    \mathbf{B} &= \mathbf{R}^{-1}\mathbf{S}\\
    \mathbf{W}\mathbf{\Lambda} &= \mathbf{BW}
\end{align}
One of the limitations of PCA is that the eigenvectors matrix is orthogonal, meaning that all components must be orthogonal. That constraint comes from the eigendecomposition on a symmetric matrix, and all covariance matrices are symmetric. This constraint is a limitation because sources that are correlated in the channel space cannot be separated by PCA (e.g., Figure \ref{fig:simpleExample}). However, the product of two symmetric matrices is generally not symmetric, and so the eigenvectors of a GED may be correlated (though they are linearly independent for distinct eigenvalues). This is the reason that sources that are correlated in channel space can be separated by GED (and will be orthogonal in the GED space). Also note that matrix $\mathbf{R}$ is inverted "on paper"; in practice, the GED is solved without explicitly inverting matrices \citep{Ghojogh2019Mar}, which means that GED can be performed on reduced-rank data.

The two matrices that GED returns contain pairs of eigenvectors and eigenvalues, with each eigenvector (spatial filter) having a corresponding eigenvalue (multivariate ratio). In this pairing, the eigenvector $\mathbf{w}_i$ points in a specific direction in the dataspace, but does not convey the importance of that direction (the $\mathbf{w}$'s are unit-length in the space of $\mathbf{R}$ because $\mathbf{W}^\text{T}\mathbf{RW}=\mathbf{I}$). In contrast, the corresponding eigenvalue $\lambda_i$ encodes the importance of the direction, but is a scalar and therefore has no intrinsic direction. The implication of this is that the eigenvector associated with the largest eigenvalue is the spatial filter that maximizes the ratio $\mathbf{S}:\mathbf{R}$ along direction $\mathbf{w}_1$. The next-largest eigenvalue is paired with the eigenvector that maximizes that ratio while being $\mathbf{R}$-orthogonal to the first direction (that is, satisfying the constraint that $\mathbf{w}_2^\text{T}\mathbf{X}$ is orthogonal to $\mathbf{w}_1^\text{T}\mathbf{X}$ \citep{parra2019correlated}). And so on for all other directions.

It is not reasonable to expect that all components are meaningful and interpretable. Indeed, if the two covariance matrices are similar to each other (e.g., created from two different experiment conditions that are matched on many perceptual and motor features), there may be only one significant component --- or possibly no significant components. The GED simply returns all solutions without a p-value or confidence interval that would indicate interpretability. Inferential statistics can be applied to the eigenvalue spectrum, which is described in section \ref{sec:stats}.

GED is easy to implement in MATLAB. The returned solutions are not guaranteed to be sorted by eigenvalue magnitude, and thus it is convenient to sort the solution matrices.
\begin{verbatim}
[W,L] = eig(S,R);
[eigvals,sidx] = sort(diag(L),'descend');
eigvecs = W(:,sidx);
\end{verbatim}

After sorting, the component time series and component map are created by multiplying an eigenvector with, respectively, the multichannel data and the covariance matrix (discussed in more detail in section \ref{sec:apply2data}).
\begin{verbatim}
comp_ts  = eigvecs(:,1)' * data; % data are chansXtime
comp_map = eigvecs(:,1)' * S;
\end{verbatim}
Trial-related data are often stored as 3D matrices (e.g., channels, time, trials), and so the code may need to be modified to reshape the data to 2D for the multiplication, and then back into 3D for further analyses.

The computation of a component time series is an important divergence from typical machine-learning classification or discriminant analyses: The goal of traditional classification analyses is to use $\mathbf{w}$ to binarize the data and predict whether the data were drawn from "class A" or "class B". However, the application of GED described here results in a continuous time series that has considerably richer information than a binary class label.

\subsection{Assumptions underlying GED}\label{sec:assumptions}
GED relies on several implicit and explicit assumptions.
\begin{enumerate}
    \item Signals mix linearly in the physical data channels. This assumption is necessary because GED is a linear decomposition. For M/EEG/LFP, this assumption is readily feasible, because electrical fields propagate instantaneously and sum linearly (within measurement capabilities) \citep{Nunez2006}. For other measurement modalities like fMRI, multiunit activity, or 2-photon imaging, the linear-mixing assumption can be interpreted that spatially distributed manifestations of the neural computations are simultaneously active (within a reasonable tolerance given by the sampling rate) and sum linearly. This assumption appears to be met in fMRI data \citep{vanDijk2020Mar,Boynton2012Aug}.
    \item The targeted features of the data are stable within the windows used to construct the covariance matrices. Stability here means that the characteristics of the signal remain consistent over the covariance matrix time window. If the underlying neural dynamics change qualitatively within that time window, for example if multiple distinct sensory/cognitive/motor processes are engaged, then the resulting GED decomposition may be less well-separated, more difficult to interpret, or less statistically robust.
    \item Covariance is a meaningful basis for source separation. Covariance is a second-order statistical mixed moment, and captures all of the pairwise linear interactions. This is notably distinct from, for example, ICA, which begins by whitening the data and optimizes based on higher-order moments such as kurtosis or entropy. Validation studies in simulated data (described later) confirm the viability of this assumption.
    \item The data features used to create the $\mathbf{S}$ and $\mathbf{R}$ matrices are physiologically or cognitively sufficiently different to produce an interpretable spatial filter. To take an extreme example, designing a spatial filter to separate 10.0 Hz from 10.1 Hz EEG activity is unlikely to produce meaningful results. The data features should be selected such that they are comparable in many aspects and differ in one or a small number of key aspects. Analogously, an experimental task condition and its associated control condition should be carefully designed to be similar but meaningfully differentiable.
\end{enumerate}

A final, implicit, assumption of GED is that the components "look reasonable" to our intuition. Our training and experiences as neuroscientists have instilled in us a sense of what data "should" look like, in terms of spatial and temporal characteristics (e.g., smoothness or spectral energy distributions) and statistical effects sizes. Data or results that are inconsistent with our expectations are likely to be rejected, or the analysis redone using different parameters or data features. This is not unique to GED --- it is nearly ubiquitous in science that our expectations, intuitions, and experiences leads us to evaluate data and results in certain ways. This is a subtle but important aspect of high-quality research: On the one hand, artifacts, noise, and confounds can be detected by experienced scientists while being missed by novices. But on the other hand, we must take care to avoid overfitting, researcher degrees of freedom, and p-hacking. 

Careful experiment and analysis design, along with appropriate inferential statistics, can help ensure that results are valid and generalizable. But it is also important to be cognizant of the role --- and potential bias --- that researcher expectations can have on data analysis pipelines. The implications of researcher overfitting are discussed extensively elsewhere \citep{Ioannidis2008Sep,Head2015Mar,Kriegeskorte2009May}.

\subsection{Understanding and avoiding overfitting}
Overfitting is a term in statistics and machine learning that refers to models being optimized for a desired result, at the expense of generalizability. Overfitting is potentially dangerous because the model parameters can be driven by noise or other non-reproducible patterns in the data.

On the other hand, overfitting is a powerful and useful approach when used appropriately. GED is based on fitting a spatial filter that maximizes the contrast  between matrices $\mathbf{S}$ and $\mathbf{R}$. Thus, one can use GED to leverage overfitting in a beneficial way without introducing systematic biases into the analyses that could confound the results. This requires some additional considerations, compared to blind decomposition methods such as ICA or PCA (these methods are "blind" in that they do not involve specifying a contrast that guides the separation).

 There are four approaches that will help evaluate whether the results reflect overfitting noise or sampling variability, and thus whether the model is likely to generalize to new data.
\begin{enumerate}
    \item Apply statistical contrasts that are orthogonal to the maximization criteria. For example, use GED to create a spatial filter that maximizes gamma-band activity across all experiment conditions, and then statistically compare gamma activity between conditions. In this case, a statistical test of the presence of gamma activity \textit{per se} is biased by the creation of the spatial filter. But a statistical test on condition differences in gamma is not biased by the filter construction. Note that this method does not guarantee an unbiased test in all cases. For example, a trial count imbalance could bias the GED towards the condition that contributes the most data.
    \item Use cross-validation to evaluate generalization performance. Cross-validation is a commonly used method in machine learning, and involves fitting the model (the spatial filter) using part of the data and applying the model to a small portion of the data that was not used to train the model. Typical train/test splits are 80/20\% or 90/10\%. One cross-validation fold would mean including only 10\% of the experimental data, which is unlikely to provide a sufficient amount of data in typical electrophysiology experiments. An alternative is to use $k$-fold cross-validation, where the analysis is repeated, each time using a different 90/10 split of the data. Thus, after 10 splits, all of the data are included, without any of the "test" trials included in the GED construction.
    \item Create the spatial filter based on independent data. This is similar to cross-validation and is a technique that is sometimes used in fMRI localizer tasks (for example, having a separate experiment acquisition to identify the fusiform face area). Thus, the GED spatial filter would be created from a different set of trials, possibly in a different experiment block.
    \item Apply inferential statistics (via permutation-testing) to evaluate the probability that a component would arise given overfitting of data when the null hypothesis is true. This is because overfitting noise will produce a maximum eigenvalue greater than 1 (the H$_0$ value), but that eigenvalue is unlikely to survive a significance test with an appropriate p-value. The mechanism for this is described below.
\end{enumerate}

\subsection{Inferential statistics} \label{sec:stats}
As mentioned earlier, the eigenvectors simply point in a particular direction in the data space, while the eigenvalues encode the importance of that direction for separating $\mathbf{S}$ from $\mathbf{R}$. Thus, the goal of inferential statistics of GED is to determine the probability that eigenvalues as large as observed could have been obtained by chance (due to noise and sampling variability), given that the null hypothesis is true. The expected value of $\lambda$ is 1 when the two covariance matrices are the same; but in real data, $\mathbf{S}$ and $\mathbf{R}$ can be expected to differ due to sampling variability and noise, even if they are drawn from the same population. Thus, the eigenvalues can be expected to be \textit{distributed} around 1, and we can expect that roughly one-half of the eigenvalues of a GED on null-hypothesis data will be larger than 1.

Inferential statistical evaluation of GED solutions involves creating an empirical null-hypothesis distribution of eigenvalues, and comparing the empirical eigenvalue relative to that distribution. One might initially think of generating covariance matrices from random numbers. However, this approach would produce an inappropriately liberal statistical threshold, because real data have a considerable amount of spatiotemporal structure that must be incorporated into the null hypothesis distribution \citep{Theiler1992Sep}. Thus, an appropriate approach is to randomize the mapping of data into covariance matrix, without generating fake data.

Consider an experiment with 100 trials, and the GED is based on the pre-trial to post-trial contrast. Each trial provides two covariance matrices, and thus there are 200 covariance matrices in total. To generate a null-hypothesis GED, each covariance matrix is randomly assigned to be averaged into $\mathbf{S}$ or $\mathbf{R}$, and a GED is performed. From the resulting collection of $M$ eigenvalues, the largest is selected. This is the largest eigenvalue that arose under the null hypothesis. This shuffling procedure is then repeated many times (e.g., 1000), with each iteration having a new random assignment of data segment to covariance matrix. The resulting collection of 1000 pseudo-$\lambda$'s creates a distribution of the largest eigenvalues expected under the null hypothesis that $\mathbf{S}=\mathbf{R}$. A p-value can be computed as the proportion of pseudo-$\lambda$ values that are larger than $\lambda$. For example, if 4/1000 null-hypothesis maximum eigenvalues were larger than the observed $\lambda$, the p-value associated with that GED component would be p=.004. Note that these tests are one-tailed, as we are interested only in components that are larger than the null-hypothesis distribution. Using the largest eigenvalue from each null-hypothesis GED corrects for multiple comparisons over the $M$ eigenvalues. This procedure is called extreme-value correction or maxT correction \citep{Nichols2002Jan,Maris2007Aug}.

Alternatively, one can define a critical threshold above which any observed $\lambda$ can be considered statistically significant, as the 95\% percentile of the pseudo-$\lambda$ distribution. If that numerical value is, e.g., 1.7, then any GED component with an associated $\lambda>1.7$ can be considered statistically significant at a 5\% threshold.

Unfortunately, things are not always so simple. If $\mathbf{S}$ and $\mathbf{R}$ are not in the same scale, then the expected $\lambda$ under the null hypothesis might be different from 1. This can occur, for example, if the $\mathbf{S}$ matrix comes from narrowband filtered data while the $\mathbf{R}$ matrix comes from broadband data. A narrowband signal will contain less variance than the broadband signal from which it was extracted. Randomly shuffling data segments into $\mathbf{S}$ or $\mathbf{R}$ might produce a pseudo-$\lambda$ distribution around 1, whereas the empirical largest $\lambda$ might be .1. If overall differences in covariance matrix magnitude are expected, the matrices can be normalized. Normalization of the covariance matrices is discussed in section \ref{sec:computingTheCovs}.

The methods described above are for statistical evaluation of a GED within one individual. But neuroscience studies typically involve group-level analyses, in which the goal is to determine whether the features of the data are consistent across individuals, which provides support for generalizability to a population. In many cases, it is advantageous to take one component per subject (e.g., the component that maximizes the desired $\mathbf{S:R}$ contrast; component selection is discussed in section \ref{sec:whichComp}), and $N$ subjects would therefore produce $N$ components. Group-level analyses would then be done on the results of the component time series analyses: ERP, power spectrum, time-frequency power, or whatever is the relevant result. Thus, procedures for group-level analyses are the same as those for analyzing data from one electrode per subject.

\section{Practical aspects}

\subsection{Preparing data for GED}
GED doesn't "know" what is real brain signal and what is noise or artifact; it simply identifies the patterns in the data that maximally separate two covariance matrices. Therefore, the data should be properly cleaned prior to GED. This may include rejecting noisy trials, temporal filtering, ICA for projecting out non-brain sources, and excluding bad channels. It is also possible to clean the covariance matrices by excluding any segment covariances that are "far away" from their mean (discussed in section \ref{sec:computingTheCovs}). Data cleaning procedures are often equipment- and lab-specific, and so data cleaning for GED can follow the same data cleaning protocol already established in the lab. GED can produce good solutions with reduced-rank data (discussed in section \ref{sec:regu}), so there is no concern about removing artifact ICs.

Channel interpolation is not necessary, because the interpolated channels are linear combinations of other channels, and thus do not contribute unique information to the decomposition. Removed channels can be interpolated in the component maps to facilitate averaging across subjects.

For EEG and LFP, the reference scheme does not impact the GED solution. Re-referencing is a linear operation, and GED is a linear decomposition. Of course, the channel weights will differ if a different reference is used,  and thus the filter forward model will also change. Referencing is an oft-debated issue in EEG research \citep{Yao2019Jul}, and average reference is often recommended. The only real restriction on re-referencing with GED is that the same reference must be used to compute the GED and the components. For example, one should not construct the GED using earlobe reference and then apply the spatial filter to average-referenced data.

\subsection{Selecting data features for \textbf{S} and \textbf{R} covariance matrices}
Selecting two data features for the GED to separate is the single most important decision that the researcher makes during a GED-based data analysis. It is also the reason why GED is such a flexible and versatile analytic backbone \citep{de2014joint,Parra2005Nov}. This means that GED forces researchers to think carefully and critically about their hypotheses and analyses, which is likely to have positive effects on the quality of the research.

The main hard constraint is that the data channels must be the same and in the same order; it is not possible to separate covariance matrices of different sizes, nor is a GED on covariance matrices with channels in different orders interpretable.

There are many features of the data to select for the contrast-enhancing ($\mathbf{S}$) and reference ($\mathbf{R}$) matrices; below is a non-exhaustive list that illustrates the possibilities.

\begin{enumerate}
    \item \textbf{Condition differences}. The data for the $\mathbf{S}$ covariance matrix come from the condition of interest (e.g., trials with an informative attentional cue) and the data for the $\mathbf{R}$ matrix come from a control condition (e.g., trials with an uninformative attentional cue) \citep{zuure2021narrowband}. One must be mindful that experiment confounds might bias the GED result. For example, if condition $\mathbf{S}$ has faster reaction times than condition $\mathbf{R}$, then the GED result might reflect motor rather than attentional processing.
    \item \textbf{Task effects}. The data for the $\mathbf{S}$ covariance matrix come from a within-trial time window (e.g., 0 to 800 ms post-trial-onset) and the data for the $\mathbf{R}$ matrix come from a pre-trial baseline period (e.g., -500 to 0 ms) \citep{duprez2020midfrontal}. When data from all conditions are pooled together, this approach avoids the risk of overfitting to one condition.
    \item \textbf{Spectral contrast}. The data for the $\mathbf{S}$ matrix are narrowband filtered in some range of interest (e.g., alpha, ~10 Hz), and the data for the $\mathbf{R}$ matrix are broadband \citep{de2015scanning} or narrowband from neighboring frequencies \citep{nikulin2011novel}.
    \item \textbf{Trial average}. The $\mathbf{S}$ matrix is computed from the trial-averaged response such as an ERP, and the $\mathbf{R}$ matrix is computed from the average of the single-trial covariance matrices \citep{Rivet2009Aug,Tanaka2019Aug,Dmochowski2015Apr}. This approach will enhance the phase-locked features of the signal.
\end{enumerate}

\subsection{Computing the covariance matrices} \label{sec:computingTheCovs}
The quality of the GED result depends entirely on the quality of the covariance matrices, so it is important to make sure that the covariance matrices are made from a sufficient amount of clean data.

One way to increase the stability of a covariance matrix is to increase the number of time points in the data segment. However, the size of the time window presents a trade-off between cognitive specificity vs. covariance quality: Shorter time windows (e.g., 100 ms) better isolate phasic sensory/cognitive/motor events, but risk a noisier covariance matrix. On the other hand, longer time windows (e.g., 1000 ms) may increase the quality of the covariance matrix but may span multiple distinct task events associated with distinct neural/cognitive systems. Keep in mind that the temporal resolution of the component time series is that of the data, and is not limited by the time windows used to create the covariance matrices. This is because the spatial filter can be applied to the entire time series data, not only the data within the covariance time window.

If the covariance matrix is computed from temporally narrowband filtered data, then the time windows to compute the covariance matrices should be at least one cycle, and preferably longer. For example, if the channel data are filtered at 4 Hz, then the time window to compute the covariance matrix should be at least 250 ms. To avoid edge effects, the temporal filter should be applied to the epoched or continuous data, not only to the data inside the covariance time window.

Ideally, the data for the two covariance matrices are of comparable quality. Matching the amount of data used for both covariance matrices (that is, the same number of trials or time points) can help ensure equal data quality. However, simply matching the amount of data does not guarantee matching covariance quality if the data used to create $\mathbf{R}$ are noisier than the data used to create $\mathbf{S}$.

When working with data that are segmented into multiple trials, the covariance matrix of each trial can be computed and then averaged over trials. Covariance-averaging can also be used for continuous (e.g., resting-state) data that are epoched into, e.g., 2-second epochs \citep{Allen2004Oct}. In this case, each individual data segment must be mean-centered before its covariance is computed. Illustrative MATLAB code shows how this can be implemented (\texttt{data} is a variable with dimensions: channel, time, trials).
\begin{verbatim}
covmat = zeros(nbchans);
for triali=1:ntrials
    seg = data(:,:,triali); % extract one trial
    seg = seg - mean(seg,2); % mean-center
    covmat = covmat + seg*seg'/(size(seg,2)-1);
end
covmat = covmat / triali;
\end{verbatim}

Computing $N$ covariance matrices also allows for an additional data cleaning step based on distance. In particular: The average covariance matrix $\overline{\mathbf{S}}$ is computed, and then the Euclidean distance (or any other distance metric) between each segment's covariance matrix $\mathbf{S}_n$ and $\overline{\mathbf{S}}$ is computed. Those $N$ distances can be z-scored, any covariance matrices with an excessive distance (e.g., $z>2.31$ corresponding to $p<.01$) are excluded, and a new average $\overline{\mathbf{S}}$ is re-computed. The justification of this procedure is that large-distance covariance segments are multivariate outliers that may bias the results.

Because covariance matrices retain the scale of the data, normalization is not necessary. Even if $\mathbf{S}$ and $\mathbf{R}$ are in different scales, normalization may be unnecessary. This is because in most cases, the relative eigenvalues are important (e.g., for sorting), not the absolute numerical values. Normalization is necessary only when (1) permutation testing is used while $\mathbf{S}$ and $\mathbf{R}$ are in different scales, or (2) data are combined from different modalities that have very different numerical ranges (e.g., EEG and MEG).

It is possible to z-normalize each data channel, however, this will change the covariance matrix, and therefore will make the filter forward model difficult to interpret, because z-normalizing each channel separately alters the magnitude of the between-channel covariances. In other words, channels with low variance are inflated whereas channels with high variance are deflated. Of course, this is the goal of z-normalizing, but some aspect of channel differences in variance are meaningful, for example alpha-band variance is higher at posterior-central channels compared to lateral temporal channels.

An alternative is to mean-center each channel separately, and then divide all channels by their pooled standard deviation. This approach preserves the relative covariance magnitudes within modality, while simultaneously ensuring that the total dataset has a pooled standard deviation of 1.

Covariance matrices are typically computed via equation \ref{eq:covariance} ($\mathbf{XX}^\text{T}$). However, alternative covariance estimators have been proposed to address certain situations, such as excessive noise or outliers. Such methods include median absolute deviation and Riemannian geometric averaging \citep{Miah2020Sep,Blum2019,Barachant2012Apr}. Covariance matrices can also be constructed using time points as features instead of, or in addition to, channels. Such time-delay embedding would turn the GED into a spatiotemporal filter, which is not discussed here. Rigorous comparisons of these estimators is beyond the scope of this paper.

\subsection{Regularization} \label{sec:regu}
Regularization involves adding some constant to the cost function of an optimization algorithm. Regularization has several benefits in machine learning, including "smoothing" the solution to reduce overfitting and increasing numerical stability, particularly for reduced-rank datasets. Regularization is not necessary for GED, but can improve its performance when there are numerical stability issues.

There are several forms of regularization, including L1 (a.k.a. Lasso), L2 (ridge), Tikhonov, shrinkage, and others. Although there are mathematical differences between different regularization techniques, it is often the case that various methods produce comparable benefits \citep[e.g.,][]{Wong2018Aug}.

Here I focus on shrinkage regularization, because it is simple and effective, and commonly used in GED applications \citep{lotte2010regularizing}. Shrinkage regularization involves adding a small number to the diagonal of the $\mathbf{R}$ matrix (and, thus, $\widetilde{\mathbf{R}}$ replaces $\mathbf{R}$ in equation \ref{eq:GEDmatrices}). That small number is some fraction of the average of $\mathbf{R}$'s eigenvalues.
\begin{align} \label{eq:regu}
    \widetilde{\mathbf{R}} &= \mathbf{R}(1-\gamma) + \gamma\alpha\mathbf{I}_M\\
    \alpha &= \sum_{i=1}^M \lambda_i
\end{align}
$\mathbf{I}_M$ is the $M\times M$ identity matrix, $\alpha$ is the average of all eigenvalues of $\mathbf{R}$, and $\gamma$ is the regularization amount. Scaling down the $\mathbf{R}$ matrix by 1-$\gamma$ ensures that $\widetilde{\mathbf{R}}$ and $\mathbf{R}$ have the same trace. This is useful because the trace of a matrix equals the sum of its eigenvalues, and thus the total "energy" of the eigenspectrum is preserved before and after regularization.

The MATLAB code is a direct implementation of equation \ref{eq:regu}.
\begin{verbatim}
gamma = .01; % 1% regularization
Rr = R*(1-gamma) + gamma*mean(eig(R))*eye(length(R));
\end{verbatim}

The effect of shrinkage regularization on GED can be understood by noting that when $\gamma=1$, the GED turns into a PCA on covariance matrix $\mathbf{S}$. Thus, regularization biases the GED towards high-variance solutions at the potential cost of reduced separability between $\mathbf{S}$ and $\mathbf{R}$. Because maximizing variance does not necessarily maximize relevance, excessive regularization may lead to results that are numerically stable but that are less sensitive to the desired contrast. For this reason, one should use as little regularization as possible but as much as necessary.

There are two ways to determine the amount of regularization to use. One is to pre-select a value for $\gamma$ and apply that value to all datasets. In our experience, for example, $\gamma=1\%$ improves the solution for matrices that are reduced-rank, noisy, or are difficult to separate, while having little or no noticeable effect on the GED solution for clean and easily separable matrices. A fixed level of regularization is not guaranteed to be an optimal amount for all datasets, but it has the advantage of being simple, unbiased, and applied equally to all of the data.

A second approach is to use cross-validation to empirically identify a $\gamma$ value \citep{parra2019correlated}. The idea is to re-compute the GED multiple times using a range of $\gamma$ values, and select the amount of shrinkage that maximizes a data feature or experiment effect of interest (e.g., the ratio of narrowband to broadband power). A potential risk of cross-validation is that the $\gamma$ parameter that maximizes one data feature (e.g., narrowband power) may reduce sensitivity to a different data feature (e.g., condition difference).

\subsection{Sign uncertainty}
Eigenvectors have a fundamental sign uncertainty. That is, because eigenvectors point along a 1D subspace (which is an infinitely long line that passes through the origin of the eigenspace), eigenvector $\mathbf{w}$ is the same as $-\mathbf{w}$. Sign uncertainties do not affect spectral or time-frequency analyses, but they do affect time-domain (ERP) analyses and topographical maps. This can cause interpretational and averaging difficulties, because, for example, a P3 ERP component can manifest as a negative deflection.

There are two principled methods for adjusting the sign of the eigenvector. One is to ensure that the electrode with the largest absolute value in the component map is positive. Thus, if the largest-magnitude electrode has a negative value, the entire eigenvector is multiplied by -1. (This method could be adapted to the ERP, for example ensuring that the component ERP deflection at 300 ms is positive.)

A second method is to assume that the eigenvector sign will be consistent for a majority of subjects, and then flip the signs of the subjects with the "minority sign." A group-averaged ERP or topographical map is computed (without any sign-flipping), and each individual subject's data are correlated with the group average. The eigenvector from subjects with a negative correlation to the group-average can be flipped, and the ERP or topographical map can be recomputed. This second method facilitates group-level data aggregation, but one must be cautious to avoid a biased statistical result if the group-level polarity is tested against zero.

\subsection{Applying the spatial filter to the data} \label{sec:apply2data}
Each eigenvector (each column of the $\mathbf{W}$ matrix) is a set of weights for computing the weighted average of all data channel time series. This is the reason that the eigenvector is called a spatial filter. Each vector-data multiplication reduces the dimensionality of the data from $M\times T$ to $1\times T$ (for $M$ channels and $T$ time points). The filter must be applied to data with the same channels in the same order as was used to create the covariance matrices.

The spatial filter is created from one feature of the data, but can be applied to other data. For example, a spatial filter created by contrasting 10 Hz activity ($\mathbf{S}$ matrix) to broadband activity ($\mathbf{R}$ matrix) can be applied to the broadband signal (this is done in a validation simulation shown later). In this case, the component is designed to optimize alpha-band activity, but it is not constrained to pass through only alpha-band activity; activity in lower or higher frequencies may also be observed in the component time series. The interpretation of, for example, 40 Hz energy in the alpha-band filter would be that the higher-frequency activity, though spectrally separate from alpha, has a similar topography as alpha, and thus passes through the alpha-optimized spatial filter.

Applying the spatial filter to data beyond what was used to construct the $\mathbf{S}$ matrix also reduces the potential for bias or overfitting.

The component time series, computed as $\mathbf{w}^\text{T}\mathbf{X}$, does not necessarily have the same units as the data $\mathbf{X}$ (e.g., $\mu V$ or fT). This is because eigenvectors have arbitrary scaling. For some applications, the units don't matter. Indeed, units do not affect qualitative or statistical comparisons across conditions or over time. Furthermore, spectral and time-frequency analyses often involve normalization to decibel or percent change, meaning that the original units don't matter. Finally, measures of phase dynamics, synchronization, and correlation are unitless, and thus data scaling has no effect.

Nonetheless, it may be useful to scale the component time series. This can be done by z-normalizing the component time series, or unit-norming the eigenvector. The formula and MATLAB code for vector normalization are shown below.
\begin{equation}
    \widetilde{\mathbf{w}} = \mathbf{w}/\|\mathbf{w}\|
\end{equation}
\begin{verbatim}
w = evecs(:,1); % first eigenvector
w = w/norm(w); % scale to unit norm
\end{verbatim}

In addition to the component time series, GED provides component maps that can be visually interpreted and averaged across individuals. These maps, however, are not the eigenvectors themselves; indeed, the eigenvectors are generally not interpretable in a physiological sense, because they simultaneous encode weights to boost channels that maximize the contrast, and suppress channels that are irrelevant or noisy \citep{haufe2014interpretation}. Instead, the component maps are obtained from the columns of $\mathbf{W}^\text{-T}$. For GED, this can be simplified without the inverse as the covariance between the component time series and the channel data ($(\mathbf{w}^\text{T}\mathbf{X})\mathbf{X}^\text{T}$) \citep{Parra2005Nov}, or as the eigenvector times the covariance matrix ($\mathbf{w}^\text{T}\mathbf{S}$) \citep{haufe2014interpretation} (a scaling factor is omitted for simplicity, as discussed above). This latter form provides a nice parallel of using the eigenvector to compute the component time series ($\mathbf{w}^\text{T}\mathbf{X}$) and the component map ($\mathbf{w}^\text{T}\mathbf{S}$).

A conceptualization is that the eigenvectors represent how to contort the channel data from “outside” to see the underlying source, whereas the component maps represent the underlying source projecting outwards onto the electrodes. That said, although the eigenvectors are not directly physiologically interpretable, they contain rich and high-spatial-frequency information, and have been used to identify empirical frequency band boundaries based on changes in spatial correlation patterns of eigenvectors across neighboring frequencies \citep{cohen2021data}.

\subsection{Which component to use} \label{sec:whichComp}
\begin{figure}[H] 
    \centering
    \includegraphics[width=.8\textwidth]{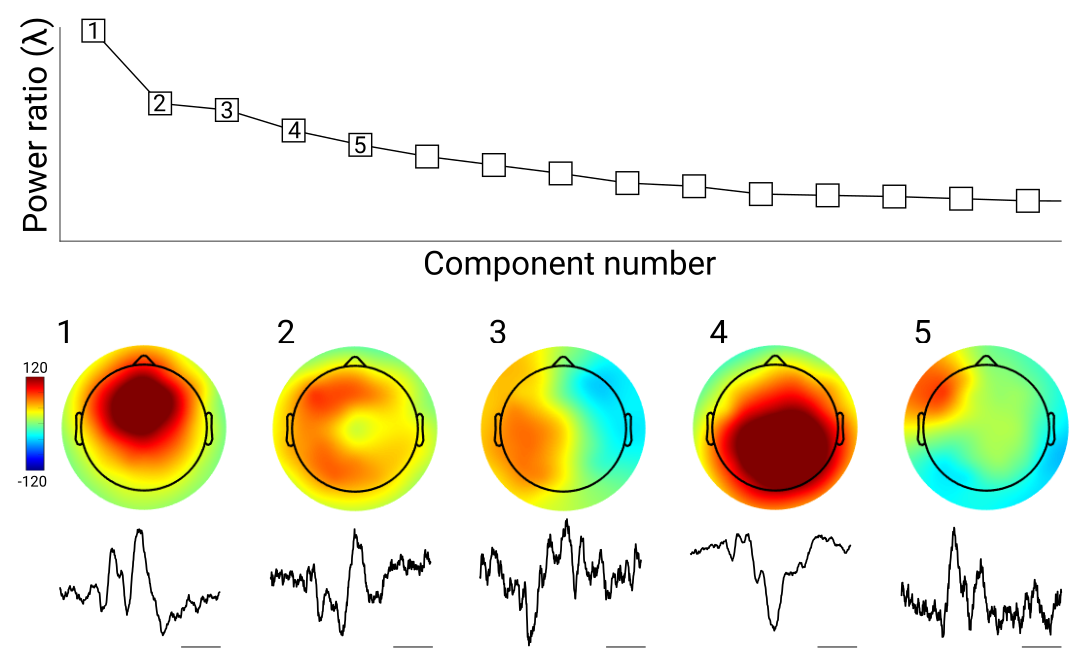}
    \caption{Picking a component based on visual inspection. A GED was performed on 64-channel EEG with the $\mathbf{S}$ matrix computed from 6-Hz filtered activity and the $\mathbf{R}$ matrix computed from broadband activity. Both matrices were created from data in a 100-800 ms window relative to trial onset. The top row shows the eigenspectum (also called a scree plot), with 15/64 components shown. The middle row shows the component maps (all maps have the same color scale shown on the left), and the bottom row shows the ERP (the spatial filter was created from narrowband data but was applied back to the broadband signal). Horizontal bar indicates 300 ms. In this example, the first component isolated midfrontal theta, while the fourth component reflects parietal theta.}\label{fig:whichcomp}
\end{figure}

Theoretically, the component with the largest eigenvalue has the best separability. However, this should be visually confirmed before applying the spatial filter to data and interpreting the results, because the component that mathematically best separates two data features is not guaranteed to be physiologically interpretable. For example, the top component in a GED that maximizes low-frequency activity may reflect eye-blink artifacts not entirely removed by ICA. In my lab, we typically produce a MATLAB figure for each dataset that shows the eigenspectrum, topoplots, and ERPs from the first 5 components (Figure \ref{fig:whichcomp}). Our default choice is the largest component, but we sometimes select later components based on topography or ERP. Picking components must be done carefully to avoid biased data selection. One possibility is to create a topographical template of the "ideal" topography (e.g., a Gaussian around electrode FCz for a midfrontal component), and then pick the GED component that has the strongest spatial correlation with that template. It is also important that any method of component selection is orthogonal to the statistical analyses performed, as discussed earlier.

Because there is theoretical motivation and mathematical justification for selecting the component with the largest eigenvalue, it is not necessary to statistically evaluate the top component, unless there is a cause for concern about overfitting. For example, if a component is created based on maximizing alpha-band activity over all conditions, the top component can be computed and the inferential statistical analyses can be done on the differences in alpha power or phase dynamics across the experiment conditions, and not on the significance of the component \textit{per se}. Analogously, in univariate analyses, one typically selects a data channel and then applies a statistical comparison across conditions, without necessarily testing whether that channel has a significant amount of activity \textit{per se}.

\subsection{Complex GED solutions}
The GED may return complex solutions, which means complex-valued eigenvalues and eigenvectors. When there are complex solutions to an eigenvalue problem, they appear in conjugate pairs. Complex solutions are often found in components with smaller eigenvalues, while the larger components have zero-valued imaginary parts. The eigenvalues (and therefore also the eigenvectors) of symmetric matrices are guaranteed to be real-valued, but $\mathbf{R^{-1}S}$ is not symmetric, and therefore its eigenvalues may be complex-valued.

Complex solutions can arise if the covariance matrices are reduced-rank or ill-conditioned, and usually indicate that the covariance matrices have poor signal-to-noise characteristics, reduced rank, or that $\mathbf{S}$ and $\mathbf{R}$ are difficult to separate. Thus, complex solutions may indicate problems with the data that should be addressed. On the other hand, reduced-rank matrices do not necessarily produce complex solutions if the data are high quality and the covariance matrices are separable.

There are several approaches to dealing with complex solutions: First, use more data to create the covariance matrices (e.g., longer time windows or wider spectral bands, or more trials); Second, redefine the GED contrast so that the matrices are more separable (e.g., all conditions against the inter-trial-interval instead of one condition against another); Third, if the data are reduced-rank, compress the data from $M$ (channels) to $r$ (matrix rank) dimensions, e.g., using PCA, and then run GED on the compressed-data covariance matrices; Fourth, apply regularization to fill in null dimensions and thus force the rank to $M$.

\subsection{Subject-specific or group-level decomposition}
If the same electrodes are placed in the same locations in different individuals, then the researcher has the choice to perform the GED separately on each individual, or to perform one GED on covariance matrices that are averaged across individuals. This is analogous to group-ICA, where data from all individuals are pooled to derive a single set of components that is based on data from all subjects \citep{Calhoun2001Nov,Calhoun2009Mar}. It is also possible to utilize GED to maximize consistency across individuals \citep{parra2019correlated}, and simultaneously within- and across-subject variance \citep{Tanaka2020Jan}.

Needless to say, the purpose of acquiring data from multiple individuals is to make inferences about a population, which motivates group-level statistical analyses. However, we must consider the features of the data we expect to be consistent across individuals, versus the features of the data we allow to vary across individuals. Performing the GED on each subject individually is based on the expectation that the statistical contrast ($\mathbf{S}$ vs. $\mathbf{R}$) is consistent across individuals, whereas the exact topography can vary due to factors such as cortical folding, functional localization, and electrode placement. Allowing for individual topographical variability also opens possibilities to explore individual differences in age, genetics, task performance, disease state, or any other subgroup analysis.

\subsection{Two-stage compression and separation}
A two-stage GED involves (1) data compression via PCA and then (2) source separation via GED. This is useful when there are many data channels, severely reduced-rank covariance matrices, or complex-valued GED solutions. The initial compression stage should be added to the analysis pipeline only when necessary, e.g., when GED returns unsatisfactory results while the data matrices are very large (e.g., hundreds of channels).

The goal of the first stage is to produce an $N\times T$ dataset, where $N$ is the number of principal components with $N<M$. This is obtained as the eigenvectors matrix $\mathbf{V}$ (of all the data, not from the GED) times the data matrix, using only the top $N$ components.
\begin{equation} \label{eq:twostage1}
    \mathbf{Y} = \mathbf{V}_{1:N}^\text{T}\mathbf{X}
\end{equation}
The subscript 1:$N$ indicates to use the first $N$ columns in the eigenvectors matrix. The number of PCA components to retain ($N$) can be based on one of two factors. First, one can use the rank of the data matrix as the dimensionality. The advantage of this approach is that it prevents information loss in the compression. In other words, the original and compressed data have the same information, but the post-PCA data matrix has fewer dimensions. This guarantees full-rank matrices in the GED and thus should improve numerical stability of the solution.

A second method is to select the number of compressed dimensions based on a statistical criterion. If the eigenvalues of the PCA are converted to percent total variance explained, then a variance threshold of, e.g., 0.1\% can be applied. The compressed signal thus comprises only principal components that contribute more than 0.1\% of the multivariate signal power. In this case, the data matrix used for GED contains less information than the original data. The motivation for this approach is that principal components that account for a tiny amount of variance might reflect noise or unrelated activity. On the other hand, the PCA is driven by total variance --- and variance does not equal relevance \citep{de2014joint}. Thus, there is a risk that some of the rejected information is important for separating $\mathbf{S}$ from $\mathbf{R}$.

The GED is then performed on the reduced-dimensional data ($\mathbf{Y}$ instead of $\mathbf{X}$). The resulting component maps, however, should be interpreted in the channel space for visualization and cross-subject averaging. Therefore, the spatial filter can be right-multiplied by the first $N$ PC vectors (the first $N$ columns in $\mathbf{V}$), which will "undo" the first-stage compression: $\mathbf{w}^\text{T}\mathbf{S}\mathbf{V}_{1:N}^\text{T}$.

\section{Validation in simulations}

We and others have done many simulations over the years to demonstrate that GED is highly accurate at reconstructing simulated activity. Some simulation protocols and results are published \citep[e.g.,][]{cohen2017comparison,Barzegaran2019Dec,Sabbagh2020Nov,nikulin2011novel}; many others are done during piloting, testing, exploring, and teaching. Simulations have the advantages of allowing full control over variables such as the amount and color of noise, the signal strength and characteristics, the number of trials, electrodes, and sources, and so on. On the other hand, simulations rarely capture the full complexity of EEG data and accompanying artifacts, and thus one should not assume that GED (or any other method) always identifies The Truth in empirical data simply because it performed well in a simulation.

Protocols for simulating EEG data range in biological and physical plausibility \citep{Naess2021Jan,Barzegaran2019Dec,Aznan2019Jan,Anzolin2021May}. For example, a pure sine wave mixed with Gaussian noise can be useful to evaluate the performance of an algorithm, but is not a neurophysiologically plausible time series.  More sophisticated models incorporate morphological and physiological details of neurons, realistic geometry, and electrical forward models \citep[e.g.,][]{Naess2021Jan}, and therefore can be used to test more precise hypotheses. The approach taken in this paper is to balance physiological plausibility (anatomical forward model, narrowband signals) with simplicity (e.g., narrowband-filtered noise to mimic a neural oscillation).

The simulation approach presented here involves generating time series signals and noise at thousands of dipole locations in the brain, passing those time series through an anatomical forward model (a.k.a. a leadfield matrix) to generate EEG activity at simulated electrodes, and then analyzing the electrode-level activity. The results can be qualitatively and quantitatively compared to the signals that were generated in the dipoles. This pipeline is useful because it is fast and efficient, allows the researcher to control the strength, nature, and locations of the signal and noise, while also providing physiologically plausible EEG topographical characteristics.

MATLAB and Python code that accompanies this tutorial are available at github.com/mikexcohen/GED\_tutorial (no additional toolboxes are required), and includes two simulations. The first is very simple (1000 ms of data, of which 500 ms is a pure sine wave) and illustrates the high reconstruction accuracy of GED and the poor performance of PCA. This provides an introduction to implementing simulated data and performing GED, and is not shown here.

\begin{figure}[H] 
    \centering
    \includegraphics[width=.9\textwidth]{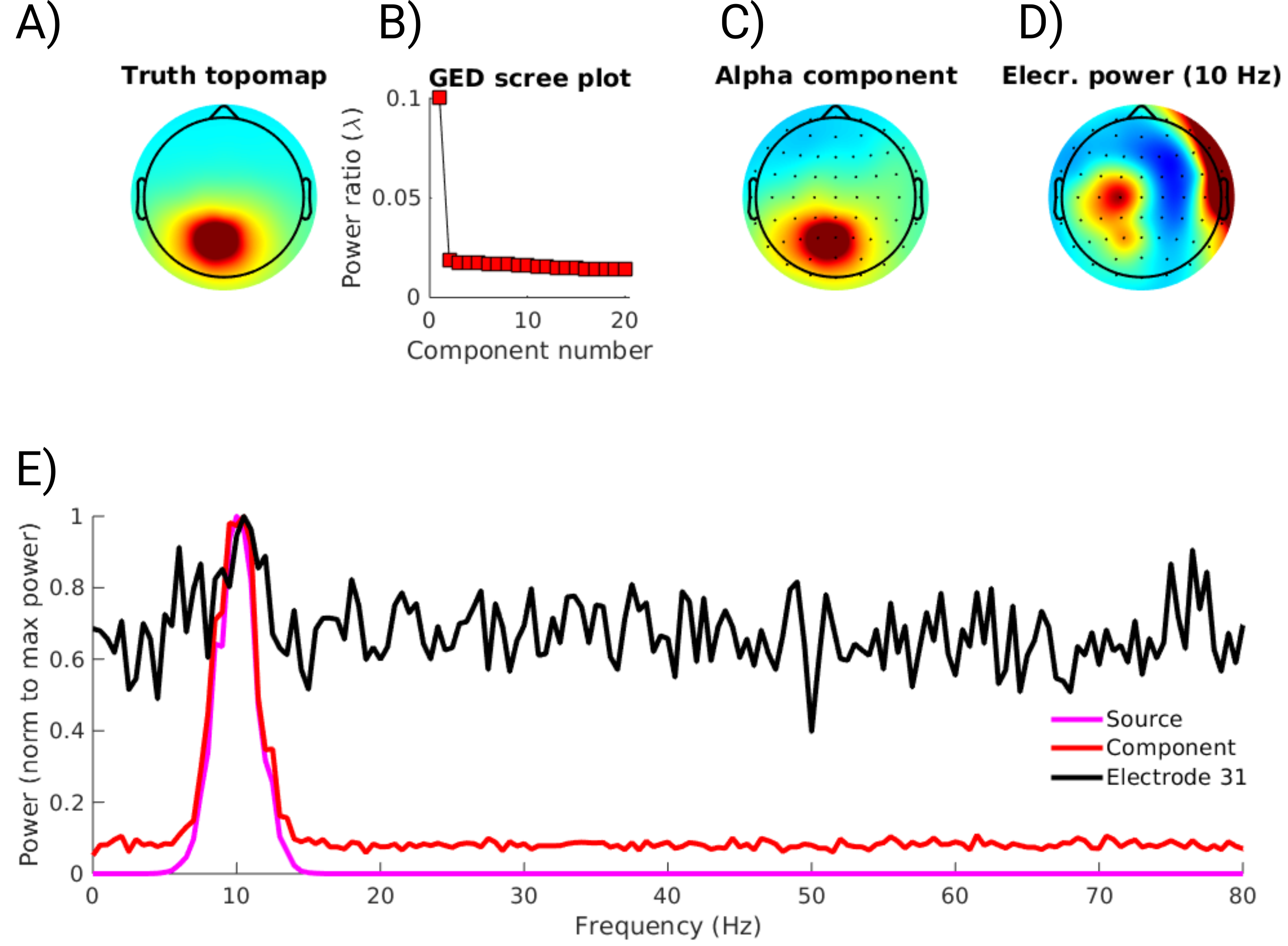}
    \caption{EEG simulation demonstrates the advantage of GED over electrode-level analyses. Data were generated in 2004 brain dipoles and projected to 64 scalp electrodes (black dots). One dipole in parietal cortex was simulated as a non-stationary alpha oscillation; all other dipoles contained Gaussian random noise with a $\sigma=1$ standard deviation. The source dipole projection is visualized in panel A. 100 segments of 2 seconds each were generated to simulate resting-state EEG data. A GED was run on 10 Hz filtered signal ($\mathbf{S}$) vs. broadband ($\mathbf{R}$). The eigenspectrum is shown in panel B. Note that all component eigenvalues are much smaller than 1, because there is more energy in the entire spectrum compared to only 10 Hz. The top component had a topography (panel C) that closely matched the ground truth. Power spectra were computed at each of 64 electrodes, and the 10 Hz power is shown in panel D. Panel E shows the power spectra from electrode 31 (closest to the ground truth dipole maximal projection), the 10 Hz GED component, and the dipole source activity (ground truth). Each topographical map has a different numerical range, so they should be interpreted qualitatively.} \label{fig:simulation}
\end{figure}

The second simulation shows a better use-case: isolating an alpha-band component embedded in noise during simulated resting-state data. Figure \ref{fig:simulation}A shows the projection of the "alpha dipole" activation onto the scalp. A GED was computed on narrowband-filtered signals ($\mathbf{S}$) compared to broadband ($\mathbf{R}$). Figure \ref{fig:simulation}B shows the scree plot, which clearly indicates one component that dominates the multivariate data. For reference, a traditional univariate analysis was applied (Welch's method on each electrode). At this SNR level, the true alpha dynamics are barely visible in the scalp and power spectrum. At higher signal gain values, the electrode data more accurately recovered the ground truth activity (not shown here, but easy to demonstrate with a minor adjustment to the code).

\section{Section 5. Interpretation}

\subsection{Does one component mean one "source"?}
As written at the outset of this paper, the term "source" has multiple interpretations, so the title of this subsection is itself an ill-posed question.

It is important to keep in mind that GED is a contrast-enhancing procedure that is based on statistical characteristics of the data. A GED component may correspond to a physiological or cognitive source, but this is not trivially guaranteed. {\color{red}Indeed, a GED component is simply a weighted sum of voltage values recorded by a collection of electrodes, which itself is not a "source."} In this sense, GED, along with other multivariate methods including ICA, can be seen as tools that, along with careful experiment design and data analyses, facilitate an attempt at source separation. Analogously, phase synchronization between brain regions does not directly indicate functional connectivity; but synchronization, along with careful experiment design and theoretical justification, can be interpreted as reflecting functional connectivity.

\begin{figure}[H] 
    \centering
    \includegraphics[width=.5\textwidth]{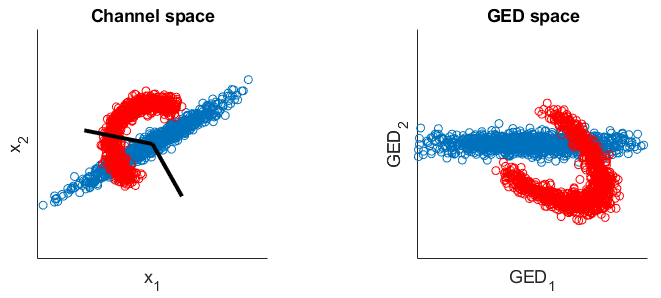}
    \caption{Because GED is a linear decomposition, nonlinear-distributed sources may not be suitably captured by GED. The black lines in the left panel show the GED eigenvectors.} \label{fig:nonlinear}
\end{figure}

Another consideration is that GED provides linear basis vectors for characterizing the data. On the one hand, this is appropriate given that electrical fields mix linearly at the electrodes. However, a linear decomposition does not mean that the data are optimally described by linear basis vectors, as illustrated in Figure \ref{fig:nonlinear}.

Another issue that impedes interpreting a component as a source is that one source can manifest in multiple components, which is discussed more below.

\subsection{Multiple significant components}
Although the component with the largest eigenvalue is theoretically the most relevant (that is, it maximally enhances the desired contrast), it is possible that multiple linearly separable dimensions separate $\mathbf{S}$ from $\mathbf{R}$ (e.g., Figure \ref{fig:overview}). For example, in a GED comparing task vs. baseline, multiple perceptual and motor operations will separate task performance from resting state. When appropriate, permutation testing can be used to determine the number of statistically significant components from a GED.

The exact interpretation of multiple significant components is not entirely clear. The simple interpretation is that they reflect distinct brain features that separate the two covariance matrices. In other words, that there is a multidimensional subspace that separates $\mathbf{S}$ and $\mathbf{R}$, and the number of significant components is the dimensionality of that subspace. Whether each dimension in this space maps onto a neural or cognitive source, however, is not guaranteed.

Furthermore, a single source can be appear in multiple components, for example, due to changes in covariance structure resulting from traveling waves, rotating dipoles, or head movements that change the locations of the EEG electrodes or placement inside the MEG. A matrix with repeated eigenvalues (or, in the case of noise, numerically distinct but statistically indistinguishable) can indicate the presence of an eigenplane. In this case, the eigenvectors associated with the repeated eigenvalues are basis vectors for that subspace, and any linear combination of the vectors is also part of the same "source."

\subsection{Do components reflect dipoles?}
GED components are a linear weighted combination of data channels, where the weights are defined by maximizing the difference between two covariance matrices. There are no constraints or assumptions about spatial smoothness or anatomical localization. This means that GED components do not necessarily correspond to a single anatomical generator that can be described by a dipole. And there is no need for them to be dipolar. This contrasts with a common view of ICA: Indeed, independent components are often evaluated in terms of their "dipolarity" (that is, their fit to a single dipole projection) \citep{Delorme2012Feb} (although ICs are not constrained to find only dipolar components). Although dipolarity may indicate physiological plausibility for sensory or motor processes that are likely to be driven by a spatially restricted population of neurons, neuroscience is increasingly moving towards the idea that many cognitive processes are implemented by spatially distributed networks \citep{Bassett2017Feb}. This suggests that the fit to a single dipole is not necessarily an important consideration when evaluating a spatial filter.

We have observed that GED components are sometimes dipole-like and other times spatially distributed (e.g., Figure \ref{fig:whichcomp}). In simulations, non-dipolar topographies can arise from noisy data, although the component time series accurately reflects the simulated ground truth. Therefore, the component topography is one indicator of the spatial filter quality, but it should not be the sole basis for selecting or rejecting a GED component. It is possible to project the component map onto the brain given a suitable forward model \citep{Cohen2017RESS,Nd2009Nov}, but this is simply a useful visualization, not a guarantee of an anatomical origin.

The recommendation here is to interpret the GED components in a conservative manner corresponding to how the components were created, which is maximizing a statistical contrast in multivariate data. The physiological interpretation is that each GED component captures a functionally cohesive and temporally coherent network, which could be spatially restricted or spatially distributed. That said, a strongly dipolar component map suggests a spatially restricted anatomical source \citep{Nunez2006}, which can facilitate a discussion of the spatial extent of the component generator.

\subsection{There is no escaping the ill-posed problem}
Finally, it is important to keep in mind that all source separation problems --- be they statistical or anatomical --- are fundamentally \textit{ill-posed}. This means that there are more unknown factors than observable measurements. Just because GED gives a solution does not mean that it gives the \textit{correct} solution, or even the only solution. There is an infinite number of possible brain generator configurations that could produce an observed EEG signal; as researchers, we must define an objective function to satisfy based on assumptions about the data.

Ultimately, the ill-posed problem means that our interpretation of GED (or any other multivariate method) should be appropriately conservative: We cannot know if we are uncovering true sources in the brain, but we can be confident that the components reflect patterns in the data that boost an hypothesis-relevant signal while also minimizing noise and individual variability that contaminate traditional univariate data selection methods.

\section{Future directions and conclusions}

\subsection{Multimodal datasets} \label{sec:multimodal}
GED does not "know" or "care" about the origin or type of data. It is possible to include multiple sources of data into the same data matrix, for example, EEG+MEG \citep{Zuure2020Sep}, LFP+single units \citep{Cohen2021Mar}, EEG+FMRI, or any other combinations of brain, behavior, and body read-outs \citep{dahne2014spoc}. Any continuous signals that are relevant for an hypothesis can be included in the covariance matrices. Multimodal datasets therefore offer a powerful means to investigate how dynamics are integrated across multiple spatial and temporal scales, and across multiple neural and behavioral observable variables.

Temporal up/downsampling may be necessarily to have data values at each time point. And data in very different scales should be suitably normalized. For example, EEG data in microvolts and MEG data in Tesla differ in raw numerical values by roughly 14 orders of magnitude.

\subsection{Better neurobiological and computational interpretations}
As a statistical method to facilitate practical data analysis, GED (and many other multivariate techniques) is unambiguously useful. Its mathematical foundations are clear, and uncountable simulation and empirical studies demonstrate its flexibility and power.

However, as neuroscientists, we are not interested in data methods \textit{per se}; we are interested in using the results of data analyses to understand brain function and its role in behavior and disease. Therefore, it would be ideal if GED components could be interpreted in a more physiological manner. It is currently premature to make claims about neural circuit configurations purely on the basis of GED topographies or eigenspectra, but there are three paths to being able to link GED to neural circuit configurations.

The first avenue is empirical multimodal studies. The idea would be to use combined EEG+fMRI, or combined EEG/LFP+spikes. One could then empirically evaluate the sensitivity of GED to distinct spatial or neural functional configurations.

A second promising avenue for linking GED to neural circuits comes from developments in computational modeling that are increasingly biophysically and morphologically accurate \citep[e.g.,][]{Naess2021Jan,Neymotin2020Jan}. Such generative forward models allow researchers to simulate data at the neural level (including specifying different types of neurons across different cortical layers, density of inter-neuronal connectivity, etc.) and then produce scalp-level MEEG data. One could then apply GED to the biophysically simulated EEG data after manipulating neurophysiologically meaningful circuit or connectivity parameters.

Finally, the component topography could be inspected for evidence of a dipolar vs. distributed pattern. Although (as written earlier), GED results should not be evaluated purely based on topographical distributions, the topography can be used to provide evidence for an anatomically restricted vs. distributed set of generators. (On the other hand, a seemingly distributed topography may also reflect noisy data.)

\subsection{Conclusions}
The goal of this paper was to present an approachable tutorial on the use of GED for multivariate source separation. There are several other lucid introductions to the theory and advantages of GED \citep{Parra2005Nov,de2014joint,haufe2014interpretation,tome2006generalized,blankertz2007optimizing}; the contribution of this paper is to focus on intuition and practical aspects that researchers might encounter while implementing GED. Although the discussion centered on electrophysiology, these methods are generic and can be applied to any type of multichannel time series data, including fMRI or calcium imaging.

Arguably, progress in neuroscience will require moving beyond traditional mass-univariate analyses, and incorporating multivariate source separation methods. GED is certainly not the only useful multivariate method, nor is it suitable for all situations and datasets. But it has many statistical and practical advantages, and its flexibility and high SNR should make it a tool in the toolbox of every neuroscientist collecting multichannel time series data.

\section*{Acknowledgements}
Many colleagues and students have helped shape my understanding and practical experience of GED. In addition to the scholars listed in the references section, following is an alphabetized list of people I have personally interacted with who indirectly contributed to this manuscript: Adam Dede, JJ Morrow, Joan Dupre, Lucas Parra, Marrit Zuure, Mihaela Gerova, Rasa Gulbinaite, Vignesh Muralidharan. I also thank three reviewers for their helpful comments.

\newpage
\bibliographystyle{elsarticle-harv}
\bibliography{cas-refs}

\begin{thebibliography}{68}
\expandafter\ifx\csname natexlab\endcsname\relax\def\natexlab#1{#1}\fi
\providecommand{\url}[1]{\texttt{#1}}
\providecommand{\href}[2]{#2}
\providecommand{\path}[1]{#1}
\providecommand{\DOIprefix}{doi:}
\providecommand{\ArXivprefix}{arXiv:}
\providecommand{\URLprefix}{URL: }
\providecommand{\Pubmedprefix}{pmid:}
\providecommand{\doi}[1]{\href{http://dx.doi.org/#1}{\path{#1}}}
\providecommand{\Pubmed}[1]{\href{pmid:#1}{\path{#1}}}
\providecommand{\bibinfo}[2]{#2}
\ifx\xfnm\relax \def\xfnm[#1]{\unskip,\space#1}\fi
\bibitem[{Ai et~al.(2018)Ai, Liu, Meng and Xie}]{Ai2018Jan}
\bibinfo{author}{Ai, Q.}, \bibinfo{author}{Liu, Q.}, \bibinfo{author}{Meng,
  W.}, \bibinfo{author}{Xie, S.Q.}, \bibinfo{year}{2018}.
\newblock \bibinfo{title}{{Chapter 6 - EEG-Based Brain Intention Recognition}},
  in: \bibinfo{booktitle}{{Advanced Rehabilitative Technology}}.
  \bibinfo{publisher}{Academic Press}, \bibinfo{address}{Cambridge, MA, USA},
  pp. \bibinfo{pages}{135--166}.
\newblock \DOIprefix\doi{10.1016/B978-0-12-814597-5.00006-0}.
\bibitem[{Allen et~al.(2004)Allen, Coan and Nazarian}]{Allen2004Oct}
\bibinfo{author}{Allen, J.J.B.}, \bibinfo{author}{Coan, J.A.},
  \bibinfo{author}{Nazarian, M.}, \bibinfo{year}{2004}.
\newblock \bibinfo{title}{{Issues and assumptions on the road from raw signals
  to metrics of frontal EEG asymmetry in emotion}}.
\newblock \bibinfo{journal}{Biol. Psychol.} \bibinfo{volume}{67},
  \bibinfo{pages}{183--218}.
\newblock \DOIprefix\doi{10.1016/j.biopsycho.2004.03.007},
  \href{http://arxiv.org/abs/15130531}{{\tt arXiv:15130531}}.
\bibitem[{Anzolin et~al.(2021)Anzolin, Toppi, Petti, Cincotti and
  Astolfi}]{Anzolin2021May}
\bibinfo{author}{Anzolin, A.}, \bibinfo{author}{Toppi, J.},
  \bibinfo{author}{Petti, M.}, \bibinfo{author}{Cincotti, F.},
  \bibinfo{author}{Astolfi, L.}, \bibinfo{year}{2021}.
\newblock \bibinfo{title}{{SEED-G: Simulated EEG Data Generator for Testing
  Connectivity Algorithms}}.
\newblock \bibinfo{journal}{Sensors} \bibinfo{volume}{21},
  \bibinfo{pages}{3632}.
\newblock \DOIprefix\doi{10.3390/s21113632}.
\bibitem[{Aznan et~al.(2019)Aznan, Atapour-Abarghouei, Bonner, Connolly,
  Moubayed and Breckon}]{Aznan2019Jan}
\bibinfo{author}{Aznan, N.K.N.}, \bibinfo{author}{Atapour-Abarghouei, A.},
  \bibinfo{author}{Bonner, S.}, \bibinfo{author}{Connolly, J.},
  \bibinfo{author}{Moubayed, N.A.}, \bibinfo{author}{Breckon, T.},
  \bibinfo{year}{2019}.
\newblock \bibinfo{title}{{Simulating Brain Signals: Creating Synthetic EEG
  Data via Neural-Based Generative Models for Improved SSVEP Classification}}.
\newblock \bibinfo{journal}{arXiv} \DOIprefix\doi{10.1109/IJCNN.2019.8852227},
  \href{http://arxiv.org/abs/1901.07429}{{\tt arXiv:1901.07429}}.
\bibitem[{Barachant et~al.(2012)Barachant, Bonnet, Congedo and
  Jutten}]{Barachant2012Apr}
\bibinfo{author}{Barachant, A.}, \bibinfo{author}{Bonnet, S.},
  \bibinfo{author}{Congedo, M.}, \bibinfo{author}{Jutten, C.},
  \bibinfo{year}{2012}.
\newblock \bibinfo{title}{{Multiclass brain-computer interface classification
  by Riemannian geometry}}.
\newblock \bibinfo{journal}{IEEE Trans. Biomed. Eng.} \bibinfo{volume}{59},
  \bibinfo{pages}{920--928}.
\newblock \DOIprefix\doi{10.1109/TBME.2011.2172210},
  \href{http://arxiv.org/abs/22010143}{{\tt arXiv:22010143}}.
\bibitem[{Barzegaran et~al.(2019)Barzegaran, Bosse, Kohler and
  Norcia}]{Barzegaran2019Dec}
\bibinfo{author}{Barzegaran, E.}, \bibinfo{author}{Bosse, S.},
  \bibinfo{author}{Kohler, P.J.}, \bibinfo{author}{Norcia, A.M.},
  \bibinfo{year}{2019}.
\newblock \bibinfo{title}{{EEGSourceSim: A framework for realistic simulation
  of EEG scalp data using MRI-based forward models and biologically plausible
  signals and noise}}.
\newblock \bibinfo{journal}{J. Neurosci. Methods} \bibinfo{volume}{328},
  \bibinfo{pages}{108377}.
\newblock \DOIprefix\doi{10.1016/j.jneumeth.2019.108377}.
\bibitem[{Bassett and Sporns(2017)}]{Bassett2017Feb}
\bibinfo{author}{Bassett, D.S.}, \bibinfo{author}{Sporns, O.},
  \bibinfo{year}{2017}.
\newblock \bibinfo{title}{{Network neuroscience}}.
\newblock \bibinfo{journal}{Nat. Neurosci.} \bibinfo{volume}{20},
  \bibinfo{pages}{353--364}.
\newblock \DOIprefix\doi{10.1038/nn.4502},
  \href{http://arxiv.org/abs/28230844}{{\tt arXiv:28230844}}.
\bibitem[{Blankertz et~al.(2007)Blankertz, Tomioka, Lemm, Kawanabe and
  Muller}]{blankertz2007optimizing}
\bibinfo{author}{Blankertz, B.}, \bibinfo{author}{Tomioka, R.},
  \bibinfo{author}{Lemm, S.}, \bibinfo{author}{Kawanabe, M.},
  \bibinfo{author}{Muller, K.R.}, \bibinfo{year}{2007}.
\newblock \bibinfo{title}{Optimizing spatial filters for robust eeg
  single-trial analysis}.
\newblock \bibinfo{journal}{IEEE Signal processing magazine}
  \bibinfo{volume}{25}, \bibinfo{pages}{41--56}.
\bibitem[{Blum et~al.(2019)Blum, Jacobsen, Bleichner and Debener}]{Blum2019}
\bibinfo{author}{Blum, S.}, \bibinfo{author}{Jacobsen, N.S.J.},
  \bibinfo{author}{Bleichner, M.G.}, \bibinfo{author}{Debener, S.},
  \bibinfo{year}{2019}.
\newblock \bibinfo{title}{{A Riemannian Modification of Artifact Subspace
  Reconstruction for EEG Artifact Handling}}.
\newblock \bibinfo{journal}{Front. Hum. Neurosci.} \bibinfo{volume}{0}.
\newblock \DOIprefix\doi{10.3389/fnhum.2019.00141}.
\bibitem[{Boynton et~al.(2012)Boynton, Engel and Heeger}]{Boynton2012Aug}
\bibinfo{author}{Boynton, G.M.}, \bibinfo{author}{Engel, S.A.},
  \bibinfo{author}{Heeger, D.J.}, \bibinfo{year}{2012}.
\newblock \bibinfo{title}{{Linear Systems Analysis of the fMRI Signal}}.
\newblock \bibinfo{journal}{Neuroimage} \bibinfo{volume}{62},
  \bibinfo{pages}{975}.
\newblock \DOIprefix\doi{10.1016/j.neuroimage.2012.01.082}.
\bibitem[{Buzs{\ifmmode\acute{a}\else\'{a}\fi}ki(2010)}]{Buzsaki2010Nov}
\bibinfo{author}{Buzs{\ifmmode\acute{a}\else\'{a}\fi}ki, G.},
  \bibinfo{year}{2010}.
\newblock \bibinfo{title}{{Neural syntax: cell assemblies, synapsembles and
  readers}}.
\newblock \bibinfo{journal}{Neuron} \bibinfo{volume}{68}, \bibinfo{pages}{362}.
\newblock \DOIprefix\doi{10.1016/j.neuron.2010.09.023}.
\bibitem[{Calhoun et~al.(2001)Calhoun, Adali, Pearlson and
  Pekar}]{Calhoun2001Nov}
\bibinfo{author}{Calhoun, V.D.}, \bibinfo{author}{Adali, T.},
  \bibinfo{author}{Pearlson, G.D.}, \bibinfo{author}{Pekar, J.J.},
  \bibinfo{year}{2001}.
\newblock \bibinfo{title}{{A method for making group inferences from functional
  MRI data using independent component analysis}}.
\newblock \bibinfo{journal}{Hum. Brain Mapp.} \bibinfo{volume}{14},
  \bibinfo{pages}{140--151}.
\newblock \DOIprefix\doi{10.1002/hbm.1048},
  \href{http://arxiv.org/abs/11559959}{{\tt arXiv:11559959}}.
\bibitem[{Calhoun et~al.(2009)Calhoun, Liu and Adali}]{Calhoun2009Mar}
\bibinfo{author}{Calhoun, V.D.}, \bibinfo{author}{Liu, J.},
  \bibinfo{author}{Adali, T.}, \bibinfo{year}{2009}.
\newblock \bibinfo{title}{{A review of group ICA for fMRI data and ICA for
  joint inference of imaging, genetic, and ERP data}}.
\newblock \bibinfo{journal}{Neuroimage} \bibinfo{volume}{45},
  \bibinfo{pages}{Suppl}.
\newblock \DOIprefix\doi{10.1016/j.neuroimage.2008.10.057},
  \href{http://arxiv.org/abs/19059344}{{\tt arXiv:19059344}}.
\bibitem[{de~Cheveign{\'e} and Arzounian(2015)}]{de2015scanning}
\bibinfo{author}{de~Cheveign{\'e}, A.}, \bibinfo{author}{Arzounian, D.},
  \bibinfo{year}{2015}.
\newblock \bibinfo{title}{Scanning for oscillations}.
\newblock \bibinfo{journal}{Journal of neural engineering}
  \bibinfo{volume}{12}, \bibinfo{pages}{066020}.
\bibitem[{de~Cheveign{\'e} and Parra(2014)}]{de2014joint}
\bibinfo{author}{de~Cheveign{\'e}, A.}, \bibinfo{author}{Parra, L.C.},
  \bibinfo{year}{2014}.
\newblock \bibinfo{title}{Joint decorrelation, a versatile tool for
  multichannel data analysis}.
\newblock \bibinfo{journal}{Neuroimage} \bibinfo{volume}{98},
  \bibinfo{pages}{487--505}.
\bibitem[{Cichy and Pantazis(2017)}]{Cichy2017Sep}
\bibinfo{author}{Cichy, R.M.}, \bibinfo{author}{Pantazis, D.},
  \bibinfo{year}{2017}.
\newblock \bibinfo{title}{{Multivariate pattern analysis of MEG and EEG: A
  comparison of representational structure in time and space}}.
\newblock \bibinfo{journal}{Neuroimage} \bibinfo{volume}{158},
  \bibinfo{pages}{441--454}.
\newblock \DOIprefix\doi{10.1016/j.neuroimage.2017.07.023},
  \href{http://arxiv.org/abs/28716718}{{\tt arXiv:28716718}}.
\bibitem[{Cohen(2017a)}]{cohen2017comparison}
\bibinfo{author}{Cohen, M.X.}, \bibinfo{year}{2017}a.
\newblock \bibinfo{title}{Comparison of linear spatial filters for identifying
  oscillatory activity in multichannel data}.
\newblock \bibinfo{journal}{Journal of neuroscience methods}
  \bibinfo{volume}{278}, \bibinfo{pages}{1--12}.
\bibitem[{Cohen(2017b)}]{cohen2017multivariate}
\bibinfo{author}{Cohen, M.X.}, \bibinfo{year}{2017}b.
\newblock \bibinfo{title}{Multivariate cross-frequency coupling via generalized
  eigendecomposition}.
\newblock \bibinfo{journal}{ELife} \bibinfo{volume}{6},
  \bibinfo{pages}{e21792}.
\bibitem[{Cohen(2021)}]{cohen2021data}
\bibinfo{author}{Cohen, M.X.}, \bibinfo{year}{2021}.
\newblock \bibinfo{title}{A data-driven method to identify frequency boundaries
  in multichannel electrophysiology data}.
\newblock \bibinfo{journal}{Journal of Neuroscience Methods}
  \bibinfo{volume}{347}, \bibinfo{pages}{108949}.
\bibitem[{Cohen et~al.(2021)Cohen, Englitz and
  Fran{\ifmmode\mbox{\c{c}}\else\c{c}\fi}a}]{Cohen2021Mar}
\bibinfo{author}{Cohen, M.X.}, \bibinfo{author}{Englitz, B.},
  \bibinfo{author}{Fran{\ifmmode\mbox{\c{c}}\else\c{c}\fi}a, A.S.C.},
  \bibinfo{year}{2021}.
\newblock \bibinfo{title}{{Large- and multi-scale networks in the rodent brain
  during novelty exploration}}.
\newblock \bibinfo{journal}{eNeuro}
  \DOIprefix\doi{10.1523/ENEURO.0494-20.2021}.
\bibitem[{Cohen and Gulbinaite(2017)}]{Cohen2017RESS}
\bibinfo{author}{Cohen, M.X.}, \bibinfo{author}{Gulbinaite, R.},
  \bibinfo{year}{2017}.
\newblock \bibinfo{title}{{Rhythmic entrainment source separation: Optimizing
  analyses of neural responses to rhythmic sensory stimulation}}.
\newblock \bibinfo{journal}{Neuroimage} \bibinfo{volume}{147},
  \bibinfo{pages}{43--56}.
\newblock \DOIprefix\doi{10.1016/j.neuroimage.2016.11.036},
  \href{http://arxiv.org/abs/27916666}{{\tt arXiv:27916666}}.
\bibitem[{D{\"a}hne et~al.(2014)D{\"a}hne, Meinecke, Haufe, H{\"o}hne,
  Tangermann, M{\"u}ller and Nikulin}]{dahne2014spoc}
\bibinfo{author}{D{\"a}hne, S.}, \bibinfo{author}{Meinecke, F.C.},
  \bibinfo{author}{Haufe, S.}, \bibinfo{author}{H{\"o}hne, J.},
  \bibinfo{author}{Tangermann, M.}, \bibinfo{author}{M{\"u}ller, K.R.},
  \bibinfo{author}{Nikulin, V.V.}, \bibinfo{year}{2014}.
\newblock \bibinfo{title}{Spoc: a novel framework for relating the amplitude of
  neuronal oscillations to behaviorally relevant parameters}.
\newblock \bibinfo{journal}{NeuroImage} \bibinfo{volume}{86},
  \bibinfo{pages}{111--122}.
\bibitem[{Dan et~al.(2020)Dan, Vandendriessche, Van~Paesschen, Weckhuysen and
  Bertrand}]{Dan2020Nov}
\bibinfo{author}{Dan, J.}, \bibinfo{author}{Vandendriessche, B.},
  \bibinfo{author}{Van~Paesschen, W.}, \bibinfo{author}{Weckhuysen, D.},
  \bibinfo{author}{Bertrand, A.}, \bibinfo{year}{2020}.
\newblock \bibinfo{title}{{Computationally-Efficient Algorithm for Real-Time
  Absence Seizure Detection in Wearable Electroencephalography}}.
\newblock \bibinfo{journal}{Int. J. Neural Syst.} \bibinfo{volume}{30},
  \bibinfo{pages}{2050035.}
\newblock \DOIprefix\doi{10.1142/S0129065720500355},
  \href{http://arxiv.org/abs/32808854}{{\tt arXiv:32808854}}.
\bibitem[{Das et~al.(2020)Das, Vanthornhout, Francart and
  Bertrand}]{Das2020Jan}
\bibinfo{author}{Das, N.}, \bibinfo{author}{Vanthornhout, J.},
  \bibinfo{author}{Francart, T.}, \bibinfo{author}{Bertrand, A.},
  \bibinfo{year}{2020}.
\newblock \bibinfo{title}{{Stimulus-aware spatial filtering for single-trial
  neural response and temporal response function estimation in high-density EEG
  with applications in auditory research}}.
\newblock \bibinfo{journal}{Neuroimage} \bibinfo{volume}{204},
  \bibinfo{pages}{116211}.
\newblock \DOIprefix\doi{10.1016/j.neuroimage.2019.116211}.
\bibitem[{Debener et~al.(2010)Debener, Thorne, Schneider and Viola}]{Debener}
\bibinfo{author}{Debener, S.}, \bibinfo{author}{Thorne, J.},
  \bibinfo{author}{Schneider, T.R.}, \bibinfo{author}{Viola, F.C.},
  \bibinfo{year}{2010}.
\newblock \bibinfo{title}{{Using ICA for the Analysis of Multi-Channel EEG
  Data}}, in: \bibinfo{booktitle}{{Simultaneous EEG and fMRI}}.
  \bibinfo{publisher}{Oxford University Press}, \bibinfo{address}{Oxford,
  England, UK}, pp. \bibinfo{pages}{121--135}.
\newblock \DOIprefix\doi{10.1093/acprof:oso/9780195372731.003.0008}.
\bibitem[{Delorme et~al.(2012)Delorme, Palmer, Onton, Oostenveld and
  Makeig}]{Delorme2012Feb}
\bibinfo{author}{Delorme, A.}, \bibinfo{author}{Palmer, J.},
  \bibinfo{author}{Onton, J.}, \bibinfo{author}{Oostenveld, R.},
  \bibinfo{author}{Makeig, S.}, \bibinfo{year}{2012}.
\newblock \bibinfo{title}{{Independent EEG Sources Are Dipolar}}.
\newblock \bibinfo{journal}{PLoS One} \bibinfo{volume}{7},
  \bibinfo{pages}{e30135}.
\newblock \DOIprefix\doi{10.1371/journal.pone.0030135}.
\bibitem[{van Dijk et~al.(2020)van Dijk, Fracasso, Petridou and
  Dumoulin}]{vanDijk2020Mar}
\bibinfo{author}{van Dijk, J.A.}, \bibinfo{author}{Fracasso, A.},
  \bibinfo{author}{Petridou, N.}, \bibinfo{author}{Dumoulin, S.O.},
  \bibinfo{year}{2020}.
\newblock \bibinfo{title}{{Linear systems analysis for laminar fMRI: Evaluating
  BOLD amplitude scaling for luminance contrast manipulations}}.
\newblock \bibinfo{journal}{Sci. Rep.} \bibinfo{volume}{10},
  \bibinfo{pages}{1--15}.
\newblock \DOIprefix\doi{10.1038/s41598-020-62165-x}.
\bibitem[{Dmochowski et~al.(2015)Dmochowski, Greaves and
  Norcia}]{Dmochowski2015Apr}
\bibinfo{author}{Dmochowski, J.P.}, \bibinfo{author}{Greaves, A.S.},
  \bibinfo{author}{Norcia, A.M.}, \bibinfo{year}{2015}.
\newblock \bibinfo{title}{{Maximally reliable spatial filtering of steady state
  visual evoked potentials}}.
\newblock \bibinfo{journal}{Neuroimage} \bibinfo{volume}{109},
  \bibinfo{pages}{63}.
\newblock \DOIprefix\doi{10.1016/j.neuroimage.2014.12.078}.
\bibitem[{Duprez et~al.(2020)Duprez, Gulbinaite and
  Cohen}]{duprez2020midfrontal}
\bibinfo{author}{Duprez, J.}, \bibinfo{author}{Gulbinaite, R.},
  \bibinfo{author}{Cohen, M.X.}, \bibinfo{year}{2020}.
\newblock \bibinfo{title}{Midfrontal theta phase coordinates behaviorally
  relevant brain computations during cognitive control}.
\newblock \bibinfo{journal}{NeuroImage} \bibinfo{volume}{207},
  \bibinfo{pages}{116340}.
\bibitem[{Ghojogh et~al.(2019)Ghojogh, Karray and Crowley}]{Ghojogh2019Mar}
\bibinfo{author}{Ghojogh, B.}, \bibinfo{author}{Karray, F.},
  \bibinfo{author}{Crowley, M.}, \bibinfo{year}{2019}.
\newblock \bibinfo{title}{{Eigenvalue and Generalized Eigenvalue Problems:
  Tutorial}}.
\newblock \bibinfo{journal}{arXiv} \URLprefix
  \url{https://arxiv.org/abs/1903.11240v1},
  \href{http://arxiv.org/abs/1903.11240}{{\tt arXiv:1903.11240}}.
\bibitem[{Haufe et~al.(2014)Haufe, Meinecke, G{\"o}rgen, D{\"a}hne, Haynes,
  Blankertz and Bie{\ss}mann}]{haufe2014interpretation}
\bibinfo{author}{Haufe, S.}, \bibinfo{author}{Meinecke, F.},
  \bibinfo{author}{G{\"o}rgen, K.}, \bibinfo{author}{D{\"a}hne, S.},
  \bibinfo{author}{Haynes, J.D.}, \bibinfo{author}{Blankertz, B.},
  \bibinfo{author}{Bie{\ss}mann, F.}, \bibinfo{year}{2014}.
\newblock \bibinfo{title}{On the interpretation of weight vectors of linear
  models in multivariate neuroimaging}.
\newblock \bibinfo{journal}{Neuroimage} \bibinfo{volume}{87},
  \bibinfo{pages}{96--110}.
\bibitem[{Head et~al.(2015)Head, Holman, Lanfear, Kahn and
  Jennions}]{Head2015Mar}
\bibinfo{author}{Head, M.L.}, \bibinfo{author}{Holman, L.},
  \bibinfo{author}{Lanfear, R.}, \bibinfo{author}{Kahn, A.T.},
  \bibinfo{author}{Jennions, M.D.}, \bibinfo{year}{2015}.
\newblock \bibinfo{title}{{The Extent and Consequences of P-Hacking in
  Science}}.
\newblock \bibinfo{journal}{PLoS Biol.} \bibinfo{volume}{13},
  \bibinfo{pages}{e1002106}.
\newblock \DOIprefix\doi{10.1371/journal.pbio.1002106}.
\bibitem[{Hebb(1949)}]{Hebb1949}
\bibinfo{author}{Hebb, D.O.}, \bibinfo{year}{1949}.
\newblock \bibinfo{title}{{The Organization of Behavior: A Neuropsychological
  Theory}}.
\newblock \bibinfo{publisher}{Wiley}, \bibinfo{address}{Chichester, England,
  UK}.
\bibitem[{Hild and Nagarajan(2009)}]{Nd2009Nov}
\bibinfo{author}{Hild, K.E.}, \bibinfo{author}{Nagarajan, S.S.},
  \bibinfo{year}{2009}.
\newblock \bibinfo{title}{{Source localization of EEG/MEG data by correlating
  columns of ICA and lead field matrices}}.
\newblock \bibinfo{journal}{IEEE Trans. Biomed. Eng.} \bibinfo{volume}{56},
  \bibinfo{pages}{2619--2626}.
\newblock \DOIprefix\doi{10.1109/TBME.2009.2028615},
  \href{http://arxiv.org/abs/19695993}{{\tt arXiv:19695993}}.
\bibitem[{Hyv{\"a}rinen(2011)}]{hyvarinen2011testing}
\bibinfo{author}{Hyv{\"a}rinen, A.}, \bibinfo{year}{2011}.
\newblock \bibinfo{title}{Testing the ica mixing matrix based on inter-subject
  or inter-session consistency}.
\newblock \bibinfo{journal}{NeuroImage} \bibinfo{volume}{58},
  \bibinfo{pages}{122--136}.
\bibitem[{Ioannidis(2008)}]{Ioannidis2008Sep}
\bibinfo{author}{Ioannidis, J.P.A.}, \bibinfo{year}{2008}.
\newblock \bibinfo{title}{{Why most discovered true associations are
  inflated}}.
\newblock \bibinfo{journal}{Epidemiology} \bibinfo{volume}{19},
  \bibinfo{pages}{640--648}.
\newblock \DOIprefix\doi{10.1097/EDE.0b013e31818131e7},
  \href{http://arxiv.org/abs/18633328}{{\tt arXiv:18633328}}.
\bibitem[{King and Dehaene(2014)}]{King2014Apr}
\bibinfo{author}{King, J.R.}, \bibinfo{author}{Dehaene, S.},
  \bibinfo{year}{2014}.
\newblock \bibinfo{title}{{Characterizing the dynamics of mental
  representations: the temporal generalization method}}.
\newblock \bibinfo{journal}{Trends in cognitive sciences} \bibinfo{volume}{18},
  \bibinfo{pages}{203}.
\newblock \DOIprefix\doi{10.1016/j.tics.2014.01.002}.
\bibitem[{Kriegeskorte et~al.(2009)Kriegeskorte, Simmons, Bellgowan and
  Baker}]{Kriegeskorte2009May}
\bibinfo{author}{Kriegeskorte, N.}, \bibinfo{author}{Simmons, W.K.},
  \bibinfo{author}{Bellgowan, P.S.F.}, \bibinfo{author}{Baker, C.I.},
  \bibinfo{year}{2009}.
\newblock \bibinfo{title}{{Circular analysis in systems neuroscience: the
  dangers of double dipping - Nature Neuroscience}}.
\newblock \bibinfo{journal}{Nat. Neurosci.} \bibinfo{volume}{12},
  \bibinfo{pages}{535--540}.
\newblock \DOIprefix\doi{10.1038/nn.2303}.
\bibitem[{Lawhern et~al.(2018)Lawhern, Solon, Waytowich, Gordon, Hung and
  Lance}]{Lawhern2018Oct}
\bibinfo{author}{Lawhern, V.J.}, \bibinfo{author}{Solon, A.J.},
  \bibinfo{author}{Waytowich, N.R.}, \bibinfo{author}{Gordon, S.M.},
  \bibinfo{author}{Hung, C.P.}, \bibinfo{author}{Lance, B.J.},
  \bibinfo{year}{2018}.
\newblock \bibinfo{title}{{EEGNet: a compact convolutional neural network for
  EEG-based brain-computer interfaces}}.
\newblock \bibinfo{journal}{J. Neural Eng.} \bibinfo{volume}{15},
  \bibinfo{pages}{056013.}
\newblock \DOIprefix\doi{10.1088/1741-2552/aace8c},
  \href{http://arxiv.org/abs/29932424}{{\tt arXiv:29932424}}.
\bibitem[{Lotte and Guan(2010)}]{lotte2010regularizing}
\bibinfo{author}{Lotte, F.}, \bibinfo{author}{Guan, C.}, \bibinfo{year}{2010}.
\newblock \bibinfo{title}{Regularizing common spatial patterns to improve bci
  designs: unified theory and new algorithms}.
\newblock \bibinfo{journal}{IEEE Transactions on biomedical Engineering}
  \bibinfo{volume}{58}, \bibinfo{pages}{355--362}.
\bibitem[{Makeig et~al.(2002)Makeig, Westerfield, Jung, Enghoff, Townsend,
  Courchesne and Sejnowski}]{Makeig2002Jan}
\bibinfo{author}{Makeig, S.}, \bibinfo{author}{Westerfield, M.},
  \bibinfo{author}{Jung, T.P.}, \bibinfo{author}{Enghoff, S.},
  \bibinfo{author}{Townsend, J.}, \bibinfo{author}{Courchesne, E.},
  \bibinfo{author}{Sejnowski, T.J.}, \bibinfo{year}{2002}.
\newblock \bibinfo{title}{{Dynamic brain sources of visual evoked responses}}.
\newblock \bibinfo{journal}{Science} \bibinfo{volume}{295},
  \bibinfo{pages}{690--694}.
\newblock \DOIprefix\doi{10.1126/science.1066168},
  \href{http://arxiv.org/abs/11809976}{{\tt arXiv:11809976}}.
\bibitem[{Maris and Oostenveld(2007)}]{Maris2007Aug}
\bibinfo{author}{Maris, E.}, \bibinfo{author}{Oostenveld, R.},
  \bibinfo{year}{2007}.
\newblock \bibinfo{title}{{Nonparametric statistical testing of EEG- and
  MEG-data}}.
\newblock \bibinfo{journal}{J. Neurosci. Methods} \bibinfo{volume}{164},
  \bibinfo{pages}{177--190}.
\newblock \DOIprefix\doi{10.1016/j.jneumeth.2007.03.024},
  \href{http://arxiv.org/abs/17517438}{{\tt arXiv:17517438}}.
\bibitem[{Miah et~al.(2020)Miah, Rahim and Shin}]{Miah2020Sep}
\bibinfo{author}{Miah, A.S.M.}, \bibinfo{author}{Rahim, M.A.},
  \bibinfo{author}{Shin, J.}, \bibinfo{year}{2020}.
\newblock \bibinfo{title}{{Motor-Imagery Classification Using Riemannian
  Geometry with Median Absolute Deviation}}.
\newblock \bibinfo{journal}{Electronics} \bibinfo{volume}{9},
  \bibinfo{pages}{1584}.
\newblock \DOIprefix\doi{10.3390/electronics9101584}.
\bibitem[{N{\ae}ss et~al.(2021)N{\ae}ss, Halnes, Hagen, Hagler, Dale, Einevoll
  and Ness}]{Naess2021Jan}
\bibinfo{author}{N{\ae}ss, S.}, \bibinfo{author}{Halnes, G.},
  \bibinfo{author}{Hagen, E.}, \bibinfo{author}{Hagler, Jr., D.J.},
  \bibinfo{author}{Dale, A.M.}, \bibinfo{author}{Einevoll, G.T.},
  \bibinfo{author}{Ness, T.V.}, \bibinfo{year}{2021}.
\newblock \bibinfo{title}{{Biophysically detailed forward modeling of the
  neural origin of EEG and MEG signals}}.
\newblock \bibinfo{journal}{Neuroimage} \bibinfo{volume}{225},
  \bibinfo{pages}{117467.}
\newblock \DOIprefix\doi{10.1016/j.neuroimage.2020.117467},
  \href{http://arxiv.org/abs/33075556}{{\tt arXiv:33075556}}.
\bibitem[{Neymotin et~al.(2020)Neymotin, Daniels, Caldwell, McDougal,
  Carnevale, Jas, Moore, Hines,
  H{\ifmmode\ddot{a}\else\"{a}\fi}m{\ifmmode\ddot{a}\else\"{a}\fi}l{\ifmmode\ddot{a}\else\"{a}\fi}inen
  and Jones}]{Neymotin2020Jan}
\bibinfo{author}{Neymotin, S.A.}, \bibinfo{author}{Daniels, D.S.},
  \bibinfo{author}{Caldwell, B.}, \bibinfo{author}{McDougal, R.A.},
  \bibinfo{author}{Carnevale, N.T.}, \bibinfo{author}{Jas, M.},
  \bibinfo{author}{Moore, C.I.}, \bibinfo{author}{Hines, M.L.},
  \bibinfo{author}{H{\ifmmode\ddot{a}\else\"{a}\fi}m{\ifmmode\ddot{a}\else\"{a}\fi}l{\ifmmode\ddot{a}\else\"{a}\fi}inen,
  M.}, \bibinfo{author}{Jones, S.R.}, \bibinfo{year}{2020}.
\newblock \bibinfo{title}{{Human Neocortical Neurosolver (HNN), a new software
  tool for interpreting the cellular and network origin of human MEG/EEG
  data}}.
\newblock \bibinfo{journal}{eLife} \DOIprefix\doi{10.7554/eLife.51214}.
\bibitem[{Nichols and Holmes(2002)}]{Nichols2002Jan}
\bibinfo{author}{Nichols, T.E.}, \bibinfo{author}{Holmes, A.P.},
  \bibinfo{year}{2002}.
\newblock \bibinfo{title}{{Nonparametric permutation tests for functional
  neuroimaging: A primer with examples}}.
\newblock \bibinfo{journal}{Hum. Brain Mapp.} \bibinfo{volume}{15},
  \bibinfo{pages}{1--25}.
\newblock \DOIprefix\doi{10.1002/hbm.1058}.
\bibitem[{Nikulin et~al.(2011)Nikulin, Nolte and Curio}]{nikulin2011novel}
\bibinfo{author}{Nikulin, V.V.}, \bibinfo{author}{Nolte, G.},
  \bibinfo{author}{Curio, G.}, \bibinfo{year}{2011}.
\newblock \bibinfo{title}{A novel method for reliable and fast extraction of
  neuronal eeg/meg oscillations on the basis of spatio-spectral decomposition}.
\newblock \bibinfo{journal}{NeuroImage} \bibinfo{volume}{55},
  \bibinfo{pages}{1528--1535}.
\bibitem[{Nunez et~al.(2006)Nunez, Nunez, Srinivasan, Press and
  Srinivasan}]{Nunez2006}
\bibinfo{author}{Nunez, P.L.}, \bibinfo{author}{Nunez, E.P.B.E.P.L.},
  \bibinfo{author}{Srinivasan, R.}, \bibinfo{author}{Press, O.U.},
  \bibinfo{author}{Srinivasan, A.P.C.S.R.}, \bibinfo{year}{2006}.
\newblock \bibinfo{title}{{Electric Fields of the Brain}}.
\newblock \bibinfo{publisher}{Oxford University Press},
  \bibinfo{address}{Oxford, England, UK}.
\bibitem[{Parra and Sajda(2003)}]{parra2003blind}
\bibinfo{author}{Parra, L.}, \bibinfo{author}{Sajda, P.}, \bibinfo{year}{2003}.
\newblock \bibinfo{title}{Blind source separation via generalized eigenvalue
  decomposition}.
\newblock \bibinfo{journal}{The Journal of Machine Learning Research}
  \bibinfo{volume}{4}, \bibinfo{pages}{1261--1269}.
\bibitem[{Parra et~al.(2019)Parra, Haufe and Dmochowski}]{parra2019correlated}
\bibinfo{author}{Parra, L.C.}, \bibinfo{author}{Haufe, S.},
  \bibinfo{author}{Dmochowski, J.P.}, \bibinfo{year}{2019}.
\newblock \bibinfo{title}{Correlated components analysis - extracting reliable
  dimensions in multivariate data}.
\newblock \href{http://arxiv.org/abs/1801.08881}{{\tt arXiv:1801.08881}}.
\bibitem[{Parra et~al.(2005)Parra, Spence, Gerson and Sajda}]{Parra2005Nov}
\bibinfo{author}{Parra, L.C.}, \bibinfo{author}{Spence, C.D.},
  \bibinfo{author}{Gerson, A.D.}, \bibinfo{author}{Sajda, P.},
  \bibinfo{year}{2005}.
\newblock \bibinfo{title}{{Recipes for the linear analysis of EEG}}.
\newblock \bibinfo{journal}{Neuroimage} \bibinfo{volume}{28},
  \bibinfo{pages}{326--341}.
\newblock \DOIprefix\doi{10.1016/j.neuroimage.2005.05.032},
  \href{http://arxiv.org/abs/16084117}{{\tt arXiv:16084117}}.
\bibitem[{Ritchie et~al.(2019)Ritchie, Kaplan and Klein}]{Ritchie2019Jun}
\bibinfo{author}{Ritchie, J.B.}, \bibinfo{author}{Kaplan, D.M.},
  \bibinfo{author}{Klein, C.}, \bibinfo{year}{2019}.
\newblock \bibinfo{title}{{Decoding the Brain: Neural Representation and the
  Limits of Multivariate Pattern Analysis in Cognitive Neuroscience}}.
\newblock \bibinfo{journal}{British J. Philos. Sci.} \bibinfo{volume}{70},
  \bibinfo{pages}{581--607}.
\newblock \DOIprefix\doi{10.1093/bjps/axx023},
  \href{http://arxiv.org/abs/31086423}{{\tt arXiv:31086423}}.
\bibitem[{Rivet et~al.(2011)Rivet, Cecotti, Souloumiac, Maby and
  Mattout}]{Rivet2011Aug}
\bibinfo{author}{Rivet, B.}, \bibinfo{author}{Cecotti, H.},
  \bibinfo{author}{Souloumiac, A.}, \bibinfo{author}{Maby, E.},
  \bibinfo{author}{Mattout, J.}, \bibinfo{year}{2011}.
\newblock \bibinfo{title}{{Theoretical analysis of xDAWN algorithm: Application
  to an efficient sensor selection in a p300 BCI}}, in:
  \bibinfo{booktitle}{{2011 19th European Signal Processing Conference}}.
  \bibinfo{publisher}{IEEE}, pp. \bibinfo{pages}{1382--1386}.
\newblock \URLprefix \url{https://ieeexplore.ieee.org/document/7073970}.
\bibitem[{Rivet and Souloumiac(2013)}]{rivet2013optimal}
\bibinfo{author}{Rivet, B.}, \bibinfo{author}{Souloumiac, A.},
  \bibinfo{year}{2013}.
\newblock \bibinfo{title}{Optimal linear spatial filters for event-related
  potentials based on a spatio-temporal model: Asymptotical performance
  analysis}.
\newblock \bibinfo{journal}{Signal Processing} \bibinfo{volume}{93},
  \bibinfo{pages}{387--398}.
\bibitem[{Rivet et~al.(2009)Rivet, Souloumiac, Attina and
  Gibert}]{Rivet2009Aug}
\bibinfo{author}{Rivet, B.}, \bibinfo{author}{Souloumiac, A.},
  \bibinfo{author}{Attina, V.}, \bibinfo{author}{Gibert, G.},
  \bibinfo{year}{2009}.
\newblock \bibinfo{title}{{xDAWN algorithm to enhance evoked potentials:
  application to brain-computer interface}}.
\newblock \bibinfo{journal}{IEEE Trans. Biomed. Eng.} \bibinfo{volume}{56},
  \bibinfo{pages}{2035--2043}.
\newblock \DOIprefix\doi{10.1109/TBME.2009.2012869},
  \href{http://arxiv.org/abs/19174332}{{\tt arXiv:19174332}}.
\bibitem[{Sabbagh et~al.(2020)Sabbagh, Ablin, Varoquaux, Gramfort and
  Engemann}]{Sabbagh2020Nov}
\bibinfo{author}{Sabbagh, D.}, \bibinfo{author}{Ablin, P.},
  \bibinfo{author}{Varoquaux, G.}, \bibinfo{author}{Gramfort, A.},
  \bibinfo{author}{Engemann, D.A.}, \bibinfo{year}{2020}.
\newblock \bibinfo{title}{{Predictive regression modeling with MEG/EEG: from
  source power to signals and cognitive states}}.
\newblock \bibinfo{journal}{Neuroimage} \bibinfo{volume}{222},
  \bibinfo{pages}{116893}.
\newblock \DOIprefix\doi{10.1016/j.neuroimage.2020.116893}.
\bibitem[{Schirrmeister et~al.(2017)Schirrmeister, Springenberg, Fiederer,
  Glasstetter, Eggensperger, Tangermann, Hutter, Burgard and
  Ball}]{Schirrmeister2017Nov}
\bibinfo{author}{Schirrmeister, R.T.}, \bibinfo{author}{Springenberg, J.T.},
  \bibinfo{author}{Fiederer, L.D.J.}, \bibinfo{author}{Glasstetter, M.},
  \bibinfo{author}{Eggensperger, K.}, \bibinfo{author}{Tangermann, M.},
  \bibinfo{author}{Hutter, F.}, \bibinfo{author}{Burgard, W.},
  \bibinfo{author}{Ball, T.}, \bibinfo{year}{2017}.
\newblock \bibinfo{title}{{Deep learning with convolutional neural networks for
  EEG decoding and visualization}}.
\newblock \bibinfo{journal}{Hum. Brain Mapp.} \bibinfo{volume}{38},
  \bibinfo{pages}{5391}.
\newblock \DOIprefix\doi{10.1002/hbm.23730}.
\bibitem[{Särelä and Valpola(2005)}]{Sarela2005Mar}
\bibinfo{author}{Särelä, J.}, \bibinfo{author}{Valpola, H.},
  \bibinfo{year}{2005}.
\newblock \bibinfo{title}{{Denoising Source Separation}}.
\newblock \bibinfo{journal}{Journal of Machine Learning Research}
  \bibinfo{volume}{6}, \bibinfo{pages}{233--272}.
\bibitem[{Tanaka(2020)}]{Tanaka2020Jan}
\bibinfo{author}{Tanaka, H.}, \bibinfo{year}{2020}.
\newblock \bibinfo{title}{{Group task-related component analysis (gTRCA): a
  multivariate method for inter-trial reproducibility and inter-subject
  similarity maximization for EEG data analysis - Scientific Reports}}.
\newblock \bibinfo{journal}{Sci. Rep.} \bibinfo{volume}{10},
  \bibinfo{pages}{1--17}.
\newblock \DOIprefix\doi{10.1038/s41598-019-56962-2}.
\bibitem[{Tanaka and Miyakoshi(2019)}]{Tanaka2019Aug}
\bibinfo{author}{Tanaka, H.}, \bibinfo{author}{Miyakoshi, M.},
  \bibinfo{year}{2019}.
\newblock \bibinfo{title}{{Cross-correlation task-related component analysis
  (xTRCA) for enhancing evoked and induced responses of event-related
  potentials}}.
\newblock \bibinfo{journal}{Neuroimage} \bibinfo{volume}{197},
  \bibinfo{pages}{177--190}.
\newblock \DOIprefix\doi{10.1016/j.neuroimage.2019.04.049},
  \href{http://arxiv.org/abs/31034968}{{\tt arXiv:31034968}}.
\bibitem[{Theiler et~al.(1992)Theiler, Eubank, Longtin, Galdrikian and
  Doyne~Farmer}]{Theiler1992Sep}
\bibinfo{author}{Theiler, J.}, \bibinfo{author}{Eubank, S.},
  \bibinfo{author}{Longtin, A.}, \bibinfo{author}{Galdrikian, B.},
  \bibinfo{author}{Doyne~Farmer, J.}, \bibinfo{year}{1992}.
\newblock \bibinfo{title}{{Testing for nonlinearity in time series: the method
  of surrogate data}}.
\newblock \bibinfo{journal}{Physica D} \bibinfo{volume}{58},
  \bibinfo{pages}{77--94}.
\newblock \DOIprefix\doi{10.1016/0167-2789(92)90102-S}.
\bibitem[{Tom{\'e}(2006)}]{tome2006generalized}
\bibinfo{author}{Tom{\'e}, A.M.}, \bibinfo{year}{2006}.
\newblock \bibinfo{title}{The generalized eigendecomposition approach to the
  blind source separation problem}.
\newblock \bibinfo{journal}{Digital Signal Processing} \bibinfo{volume}{16},
  \bibinfo{pages}{288--302}.
\bibitem[{Uhlhaas et~al.(2009)Uhlhaas, Pipa, Lima, Melloni, Neuenschwander,
  Nikoli{\ifmmode\acute{c}\else\'{c}\fi} and Singer}]{Uhlhaas2009Jul}
\bibinfo{author}{Uhlhaas, P.J.}, \bibinfo{author}{Pipa, G.},
  \bibinfo{author}{Lima, B.}, \bibinfo{author}{Melloni, L.},
  \bibinfo{author}{Neuenschwander, S.},
  \bibinfo{author}{Nikoli{\ifmmode\acute{c}\else\'{c}\fi}, D.},
  \bibinfo{author}{Singer, W.}, \bibinfo{year}{2009}.
\newblock \bibinfo{title}{{Neural synchrony in cortical networks: history,
  concept and current status}}.
\newblock \bibinfo{journal}{Front. Integr. Neurosci.} \bibinfo{volume}{3},
  \bibinfo{pages}{17.}
\newblock \DOIprefix\doi{10.3389/neuro.07.017.2009},
  \href{http://arxiv.org/abs/19668703}{{\tt arXiv:19668703}}.
\bibitem[{Wong et~al.(2018)Wong, Fuglsang, Hjortkj{\ae}r, Ceolini, Slaney and
  de~Cheveign{\ifmmode\acute{e}\else\'{e}\fi}}]{Wong2018Aug}
\bibinfo{author}{Wong, D.D.E.}, \bibinfo{author}{Fuglsang, S.A.},
  \bibinfo{author}{Hjortkj{\ae}r, J.}, \bibinfo{author}{Ceolini, E.},
  \bibinfo{author}{Slaney, M.},
  \bibinfo{author}{de~Cheveign{\ifmmode\acute{e}\else\'{e}\fi}, A.},
  \bibinfo{year}{2018}.
\newblock \bibinfo{title}{{A Comparison of Regularization Methods in Forward
  and Backward Models for Auditory Attention Decoding}}.
\newblock \bibinfo{journal}{Front. Neurosci.} \bibinfo{volume}{12}.
\newblock \DOIprefix\doi{10.3389/fnins.2018.00531}.
\bibitem[{Wouters et~al.(2018)Wouters, Kloosterman and
  Bertrand}]{Wouters2018Oct}
\bibinfo{author}{Wouters, J.}, \bibinfo{author}{Kloosterman, F.},
  \bibinfo{author}{Bertrand, A.}, \bibinfo{year}{2018}.
\newblock \bibinfo{title}{{Towards online spike sorting for high-density neural
  probes using discriminative template matching with suppression of interfering
  spikes}}.
\newblock \bibinfo{journal}{J. Neural Eng.} \bibinfo{volume}{15},
  \bibinfo{pages}{056005.}
\newblock \DOIprefix\doi{10.1088/1741-2552/aace8a},
  \href{http://arxiv.org/abs/29932426}{{\tt arXiv:29932426}}.
\bibitem[{Yao et~al.(2019)Yao, Qin, Hu, Dong, Vega and Sosa}]{Yao2019Jul}
\bibinfo{author}{Yao, D.}, \bibinfo{author}{Qin, Y.}, \bibinfo{author}{Hu, S.},
  \bibinfo{author}{Dong, L.}, \bibinfo{author}{Vega, M.L.B.},
  \bibinfo{author}{Sosa, P.A.V.}, \bibinfo{year}{2019}.
\newblock \bibinfo{title}{{Which Reference Should We Use for EEG and ERP
  practice?}}
\newblock \bibinfo{journal}{Brain Topogr.} \bibinfo{volume}{32},
  \bibinfo{pages}{530--549}.
\newblock \DOIprefix\doi{10.1007/s10548-019-00707-x}.
\bibitem[{Zuure and Cohen(2021)}]{zuure2021narrowband}
\bibinfo{author}{Zuure, M.B.}, \bibinfo{author}{Cohen, M.X.},
  \bibinfo{year}{2021}.
\newblock \bibinfo{title}{Narrowband multivariate source separation for
  semi-blind discovery of experiment contrasts}.
\newblock \bibinfo{journal}{Journal of Neuroscience Methods}
  \bibinfo{volume}{350}, \bibinfo{pages}{109063}.
\bibitem[{Zuure et~al.(2020)Zuure, Hinkley, Tiesinga, Nagarajan and
  Cohen}]{Zuure2020Sep}
\bibinfo{author}{Zuure, M.B.}, \bibinfo{author}{Hinkley, L.B.},
  \bibinfo{author}{Tiesinga, P.H.E.}, \bibinfo{author}{Nagarajan, S.S.},
  \bibinfo{author}{Cohen, M.X.}, \bibinfo{year}{2020}.
\newblock \bibinfo{title}{{Multiple Midfrontal Thetas Revealed by Source
  Separation of Simultaneous MEG and EEG}}.
\newblock \bibinfo{journal}{J. Neurosci.} \bibinfo{volume}{40},
  \bibinfo{pages}{7702--7713}.
\newblock \DOIprefix\doi{10.1523/JNEUROSCI.0321-20.2020}.

\end{thebibliography}

\end{document}